\documentclass[11pt,twoside,fleqno]{jcap}
\usepackage{epsfig}
\usepackage{amssymb}
\usepackage{fancyheda}
\usepackage{times}
\usepackage[varg]{txfonts}

\newcommand{\ftn}{\footnotesize}
\newcommand{\nsz}{\normalsize}
\newcommand{\ssz}{\scriptsize}

\newcommand{\TeV}{{\mbox{\rm TeV}}}
\newcommand{\MeV}{{\mbox{\rm MeV}}}
\newcommand{\GeV}{{\mbox{\rm GeV}}}
\newcommand{\keV}{{\mbox{\rm keV}}}
\newcommand{\vH}{{\mbox{$\bar H$}}}
\newcommand{\vHi}{{\mbox{$\bar H_{_{\rm I}}$}}}
\newcommand{\vrho}{{\mbox{$\bar\rho$}}}
\newcommand{\vn}{{\mbox{$\bar n$}}}
\newcommand{\vq}{{\mbox{$\bar q$}}}
\newcommand{\vQ}{{\mbox{$\bar Q$}}}
\newcommand{\vV}{{\mbox{$\bar V$}}}
\newcommand{\vVo}{{\mbox{$\bar V_0$}}}

\newcommand{\vTi}{{\mbox{$T_{{\rm I}}$}}}
\newcommand{\vti}{{\mbox{$\vtau_{_{\rm I}}$}}}

\newcommand{\vtns}{{\mbox{$\vtau_{_{\rm NS}}$}}}

\def\openep{\leavevmode\hbox{\normalsize$\iota$\kern-3.8pt$^$-}}
\newcommand{\vtau}{\ensuremath{\tauup}}
\newcommand{\vtauf}{\ensuremath{\tauup}}

\def\beq{\begin{equation}}
\def\eeq{\end{equation}}
\def\bea{\begin{eqnarray}}
\def\eea{\end{eqnarray}}

\newcommand{\Ti}{\ensuremath{T_{\rm I}}}
\newcommand{\Tkr}{\ensuremath{T_{\rm KR}}}
\newcommand{\Ts}{\ensuremath{T_{\rm SUSY}}}
\newcommand{\Tns}{\ensuremath{T_{\rm NS}}}
\newcommand{\Tc}{\ensuremath{T_{\rm C}}}

\newcommand{\Omax}{{\mbox{$\Omega_{\ax} h^2$}}}
\newcommand{\Omgr}{{\mbox{$\Omega_{\Gr} h^2$}}}
\newcommand{\Omx}{{\mbox{$\Omega_{\chia}h^2$}}}
\newcommand{\Omqns}{\ensuremath{\Omega_q^{\rm NS}}}

\newcommand\sigv[1]{\langle (v\sigma)_{#1}\rangle}
\newcommand\Gma[1]{\langle \Gamma_{#1}\rangle}
\newcommand\Gm[1]{\Gamma_{#1}}

\newcommand{\chia}{\ensuremath{X}}

\newcommand{\gl}{\ensuremath{\tilde{g}}}
\newcommand{\ax}{\ensuremath{\tilde{a}}}
\newcommand{\sq}{\ensuremath{\tilde{q}}}
\newcommand{\Gr}{\ensuremath{\widetilde{G}}}
\newcommand{\nequ}{\ensuremath{n^{\rm eq}}}

\newcommand\tsq[1]{|{\cal M}_{#1}|^2}
\def\tcm{\theta_{\rm\scriptscriptstyle CM}}

\newcommand{\sFref}[2]{Fig.~\ref{#1}-{\sf\small ({#2})}}
\newcommand{\sEref}[2]{Eq.~(\ref{#1}{\sf\small  {#2}})}

\begin{document}

\title{\huge \bfseries\scshape Quintessential Kination
and Thermal Production of Gravitinos and Axinos}

\author{\large \bfseries\scshape M.E. G\'omez$^{\rm a}$, S. Lola$^{\rm b}$, C.
Pallis$^{\rm a}$ and J. Rodr\'iguez-Quintero$^{\rm a}$}

\address{$^{\rm a}$ Departamento de
Fisica Aplicada,  \\ Facultad de Ciencias Experimentales, \\
Universidad de Huelva,  21071 Huelva, SPAIN}

\address{$^{\rm b}$ Department of Physics, University of Patras,
\\ Panepistimioupolis, GR-26500 Patras, GREECE}

\eads{mario.gomez@dfa.uhu.es, magda@physics.upatras.gr,
kpallis@gen.auth.gr, jose.rodriguez@dfaie.uhu.es}

\begin{abstract}{\vspace{5pt}\par

The impact of a kination-dominated phase generated by a
quintessential exponen-tial model on the thermal abundance of
gravitinos and axinos is investigated. We find that their
abundances become proportional to the transition temperature from
the kination to the radiation era; since this temperature is
significantly lower than the initial (``reheating") temperature,
the abundances decrease with respect to their values in the
standard cosmology. For values of the quintessential
energy-density parameter close to its upper bound, on the eve of
nucleosynthesis, we find the following: (i) for unstable
gravitinos, the gravitino constraint is totally evaded; (ii) If
the gravitino is stable, its thermal abundance is not sufficient
to account for the cold dark matter of the universe; (iii) the
thermal abundance of axinos can satisfy the cold dark matter
constraint for values of the initial temperature well above those
required in the standard cosmology. A novel calculation of the
axino production rate by scatterings at low temperature is also
presented.}

\end{abstract}

\keyw{Cosmology, Dark Energy, Dark Matter} 

\publishedin{{\sl J. Cosmol. Astropart. Phys.} {\bf 01}, 027
(2009)}

\maketitle

\tableofcontents \vskip0.10cm\noindent\rule\textwidth{.4pt}

\setcounter{page}{1} \pagestyle{fancyplain}


\rhead[\fancyplain{}{ \bf \thepage}]{\fancyplain{}{\sl\sc
Quintessential Kination and Thermal Production of Gravitinos and
Axinos}} \lhead[{\fancyplain{}{ \sc \leftmark}}]{\fancyplain{}{\bf
\thepage}} \cfoot{}

\section{Introduction}\label{intro}
\setcounter{equation}{0}

A plethora of recent data \cite{wmap, snae} indicates
\cite{wmapl} that the two major components of the present universe
are \emph{Cold Dark Matter} (CDM) and \emph{Dark Energy}
(DE) with density parameters \cite{wmap}
\beq \mbox{\sf\small (a)}~~\Omega_{\rm
CDM}=0.214\pm0.027~~\mbox{and}~~\mbox{\sf\small (b)}~~\Omega_{\rm
DE}=0.742\pm0.03 \label{cdmba}\eeq
at $95\%$ \emph{confidence level} (c.l.). Identifying the nature
of these two unknown substances, is one of the major
challenges in contemporary cosmo-particle theories.

The DE component can be explained by modifying the \emph{standard
cosmology} (SC) via the introduction of a slowly evolving scalar
field called quintessence \cite{early} (for reviews, see
Ref.~\cite{der}). An open possibility in this scenario is the
existence of an early \emph{kination dominated} (KD) era
\cite{kination}, where the universe is dominated by the kinetic
energy of the quintessence field; this period is an indispensable
ingredient of quintessential inflationary scenaria \cite{qinf,
dimopoulos, chung}. During this era, the expansion rate of the
universe is larger compared to its value during the usual
\emph{radiation domination} (RD) epoch. This implies that the
relic abundance of the \emph{weakly interacting massive particles}
(WIMPs) can be significantly enhanced \emph{with respect to}
(w.r.t) its value in the SC \cite{Kam, salati, prof, jcapa},
provided that they decouple from the thermal bath during the KD
era. Further phenomenological implications of this effect for the
future collider or astrophysics experiments have also been studied
\cite{kinpheno}.

WIMPs are the most natural candidates \cite{candidates} to account
for the second  major component of the present universe, the CDM.
Among them, the most popular is the lightest neutralino
\cite{goldberg, lkk} which turns out to be the \emph{lightest
supersymmetric particle} (LSP) in a sizeable fraction of the
parameter space of \emph{sypersymmetric} (SUSY) models and
therefore, stable under the assumption of the conservation of
$R$-parity. However, SUSY theories predict the existence of even
more weakly interacting massive particles, known as \emph{e}-WIMPs
\cite{ewimps} which can naturally play the role of LSP. These are
the gravitino, $\Gr$, and the axino, $\ax$ ($\Gr$ is the spin-3/2
fermionic SUSY partner of the graviton, and $\ax$ the spin-$1/2$
fermionic SUSY partner of the axion which arises in SUSY
extensions \cite{goto} of the \emph{Peccei-Quinn} (PQ) solution
\cite{pq} to the strong CP problem). As their name indicates, the
interaction rates of gravitinos and axinos are \emph{extremely}
weak, since they are respectively suppressed by the reduced Planck
scale, $m_{\rm P}=M_{\rm P}/\sqrt{8\pi}$ ($M_{\rm
P}=1.22\times10^{19}~{\rm GeV}$ being the Planck mass) and by the
axion decay constant, $f_a\sim(10^{10}-10^{12})~{\rm GeV}$ (for a
review, see Ref.~\cite{kim}).

Due to the weakness of their interactions, \emph{e}-WIMPs depart
from chemical equilibrium very early ($\Gr$ at a energy scale
close to $m_{\rm P}$ and $\tilde a$ close to $f_a$) and we expect
that their relic density (created due to this early decoupling) is
diluted by the primordial inflation. However, they can be
reproduced in the following ways: (i) in the thermal bath, through
scatterings \cite{gravitinoc, moroi1, Bolz, steffen, axino,
steffenaxino} and decays \cite{axino, small, strumia} involving
superpartners, and (ii) non-thermally \cite{gravitinont, axinont},
from the out-of-equilibrium decay of the \emph{next-to-LSP}
(NLSP). In this paper we do not consider the possible non-thermal
production of \emph{e}-WIMPs, since this mechanism is highly model
dependent (i.e., it is sensitive to the type and decay products of
the NLSP). As a consequence, we do not consider either the
out-of-equilibrium decay of the one \emph{e}-WIMP to the other (as
in the case of \cref{asaka}, where $\Gr$ is the NLSP and $\ax$ the
LSP). In all these cases, extra restrictions have to be imposed in
order not to jeopardize the success of the standard Big Bang
\emph{Nucleosynthensis} (NS).

The latter requirement  has to be satisfied also for unstable
$\Gr$. This restriction imposes  a tight upper bound on the
initial (``reheating") temperature, $\Ti$, of the universe in the
SC \cite{gravitinoc, moroi1,kohri, kohri2, oliveg}.  On the other
hand, if one of the \emph{e}-WIMPs is a stable LSP, it has to obey
the CDM constraint. In particular, its relic density
$\Omega_{\chia}h^2$ has to be confined in the region \cite{wmap}
\beq\label{cdmb} \mbox{\sf\small (a)}~0.097\lesssim
\Omega_{\chia}h^2 \lesssim 0.12~~\mbox{for}~~\mbox{\sf\small
(b)}~10~\keV\leq m_\chia\leq m_{\rm NLSP} \eeq
where $m_{\rm NLSP}$ is the mass of the NLSP. Let us note, in
passing, that the lower bound of \sEref{cdmb}{a} is valid under
the assumption that CDM is entirely composed by $X$'s and the
abundance of non-thermaly produced $X$'s is negligible. The lower
bound on $m_{\chia}$ arises from the fact that smaller $m_\chia$
cannot explain \cite{sformation} the observed early reionization
\cite{wmap}. For about $10\leq m_\chia/\keV\leq 100$, $X$'s may
constitute warm dark matter (the mass limits above are to be
considered only as indicative).

In this paper we reconsider the creation of a KD era in the
context of the exponential quintessential model \cite{wet, expo},
taking into account restrictions arising from NS, the inflationary
scale, the acceleration of the universe and the DE density
parameter. Although this model does not possess a tracker-type
solution \cite{salati, attr} in the allowed range of its
parameters, it can produce a viable present-day cosmology in
conjunction with the domination of an early KD era, for a
reasonable region of initial conditions \cite{brazil, cline,
silogi, german}. We then investigate the impact of KD on the
thermal production of \emph{e}-WIMPs,  solving the relevant
equations both numerically and semianalytically. We find that the
abundance of \emph{e}-WIMPs becomes proportional to the transition
temperature from KD to RD era, $\Tkr$, and decreases w.r.t its
value in the SC since  $\Tkr$ can be much lower than $\Ti$. In
particular, we consider two cases, depending on whether $\Tkr$ is
higher -- \emph{high $T$ regime} (HTR) -- or lower -- \emph{low
$T$ regime} (LTR) -- than a threshold $\Tc\simeq10~\TeV$, below
which the thermal production of $\ax$ via the decay of the
superpartners becomes important; for this low temperature region,
a novel formulae for the production of $\ax$'s through scatterings
is presented. It turns out that, in this part of the parameter
space, $\ax$ becomes an attractive CDM candidate within the
\emph{quintessential kination scenario} (QKS).

Modifications to the thermal production of \emph{e}-WIMPs have
also been investigated in the context of extra dimensional
theories \cite{seto, panot}, where the expansion rate of the
universe can be also enhanced w.r.t its value in the SC, due to
the presence of an extra term including the brane-tension.
However, this increase is more drastic than in the QKS. In
addition, the production of $\ax$ via scatterings  at low
temperature and via the decay \cite{axino} of the SUSY particles
has not been taken into account \cite{panot}.

We start our analysis by reviewing the basic features of the
exponential quintessential model in Sec.~\ref{sec:quint}. We then
present our numerical and semi-analytical calculations of the
thermal abundance of \emph{e}-WIMPs in Sec.~\ref{sec:boltz} and
study the parameter space allowed by several requirements for
$\Gr$ (Sec.~\ref{sec:grv}) and for $\ax$ (Sec.~\ref{sec:axn}). Our
conclusions are summarized in Sec.~\ref{sec:con}. Computational
issues on the low temperature $\ax$-production are discussed in
Appendix A.

Throughout the text, brackets are used by applying disjunctive
correspondence, natural units ($\hbar=c=k_{\rm B}=1$) are assumed,
the subscript or superscript $0$ refers to present-day values
(except in the coefficient $V_0$) and $\log~[\ln]$ stands for
logarithm with basis $10~[e]$. Moreover, we assume that the domain
wall number \cite{kim} is equal to 1.

\section{The Quintessential Exponential Model}\label{sec:quint}

In this section we review the system of equations which governs
the quintessential cosmological evolution (Sec.~\ref{Beqs}) and
the various observational restrictions that we impose
(Sec.~\ref{reqq}). We then describe the salient features of this
evolution in \Sref{Qev} and the allowed parameter space in
Sec.~\ref{ap}.

\subsection{The Quintessential Set-up}
\label{Beqs}

We assume the existence of a spatially homogeneous scalar field
$q$ (not to be confused with the deceleration parameter
${\sf\small q}$ in \Sref{reqq}), which obeys the Klein-Gordon
equation. In particular,
\beq \ddot q+3H\dot q+V_{,q}=0,~~\mbox{where}~~V=V_0 e^{-\lambda
q/m_{{\rm P}}}\label{qeq} \eeq
is the adopted potential for the $q$ field, the subscript $,q$
[dot] stands for derivative w.r.t  $q$ [the cosmic time, $t$] and
$H$ is the Hubble expansion parameter,
\begin{equation}\label{rhoqi}
H= \sqrt{\rho_q +\rho_{{\rm R}}+ \rho_{{\rm M}}}/\sqrt{3}m_{{\rm
P}}~~\mbox{with}~~\rho_q=\frac{1}{2}\dot q^2+V,\eeq
the energy density of $q$. The energy density of radiation,
$\rho_{{\rm R}}$, can be evaluated as a function of the
temperature, $T$, while the energy density of matter, $\rho_{{\rm
M}}$, with reference to its present-day value:
\beq\label{rhos} \rho_{{\rm R}}=\frac{\pi^2}{30}g_{\rho*}\
T^4~~\mbox{and}~~\rho_{{\rm M}}R^3=\rho^0_{{\rm M}}R_0^3
\end{equation}
($R$ being the scale factor of the universe). Assuming no entropy
production due to domination of $q$ or any other field, the
entropy density, $s$, satisfies the following equations:
\beq sR^3=s_{\rm p}R_{\rm
p}^3~~\mbox{where}~~s=\frac{2\pi^2}{45}g_{s*}\ T^3. \label{rs}\eeq
Here, the subscript ``p'' represents a specific reference point at
which the quantities $s$ and $R$ are evaluated and
$g_{\rho*}(T)~[g_{s*}(T)]$ is the energy [entropy] effective
number of degrees of freedom at temperature $T$. The numerical
values for these quantities are evaluated using the tables
included in {\tt micrOMEGAs} \cite{micro}. As it turns out,
$g_{\rho*}\simeq g_{s*}$ for $T>T_{\rm NS}=1~\MeV$, with values
$g_{\rho*}= 105.74$ for $T\simeq1~\TeV$ and $g_{\rho*}=228.75$ for
$T>1~\TeV$ assuming the particle content of the \emph{Minimal SUSY
Standard Model} (MSSM). Since the abundances under consideration
take their present value on the eve of NS ($T\simeq T_{\rm NS}$),
we do not insist in a distinction between $g_{\rho*}$ and $g_{s*}$
and set $g_{\rho*}=g_{s*}=g_*$. On the contrary, for $T<T_{\rm
NS}$ we obtain in general $g_{\rho*}<g_{s*}$. Taking into account
the existence of the cosmic background radiation at present, plus
three(almost) massless neutrino species, we get \cite{kolb}
$g^0_{\rho*}=3.36$ and $g^0_{s*}=3.91$.

The numerical integration of Eq.~(\ref{qeq}) is facilitated by
converting the time derivatives to derivatives w.r.t the
logarithmic time \cite{brazil, german}, which is defined as a
function of the redshift $z$:
\beq \vtau=\ln\left(R/R_0\right)=-\ln (1+z).\label{dtau} \eeq
Changing the differentiation and introducing the following
dimensionless quantities:
\beq \label{vrhos}\vrho_{{\rm M[R]}}=\rho_{{\rm M[R]}}/\rho^0_{\rm
c },~\vVo=V_0/\rho^0_{\rm c}~~\mbox{and}~~\vq=q/\sqrt{3}m_{{\rm
P}},\eeq
Eq.~(\ref{qeq}) turns out to be equivalent to the system of two
first-order equations:
\beq \vQ=\vH\vq^\prime ~~\mbox{and}~~\vH\vQ^\prime+3\vH\vQ+\bar
V_{,\bar q}=0~~\mbox{with}~~\vH^2=\vrho_q+\vrho_{{\rm
R}}+\vrho_{{\rm M}}, \label{vH} \eeq
where prime denotes (unless otherwise stated) derivative w.r.t
$\vtau$ and the following quantities have been defined:
\beq \vV=\vVo e^{-\sqrt{3}\lambda
\vq},~~\vH=H/H_0,~~\vQ=Q/\sqrt{\rho^0_{\rm
c}}~~\mbox{and}~~\vrho_q=\vQ^2/2+\vV.\label{vrhoq}\eeq
In our numerical calculation, we use the following values:
\beq \rho^0_{\rm c}=8.099\times10^{-47}h^2~{\rm GeV^4
}~~\mbox{and}~~H_0=2.13\times10^{-42}h~{\rm GeV}\eeq
with $h=0.72$. In addition, $\vrho^0_{{\rm M}}=0.26$ and
$T_0=2.35\times 10^{-13}~{\rm GeV}$ and from Eq.~(\ref{rhos}), we
get $\vrho^0_{{\rm R}}=8.04\times10^{-5}$.

Eq. (\ref{vH}) can be resolved numerically by specifying two
initial conditions at a logarithmic time, $\vti$, which
corresponds to a temperature $\Ti$ defined as the maximal $T$
after the end of primordial inflation, assuming instantaneous
reheating. We take $q(\vti)=0$ throughout our analysis, without
any lose of generality \cite{jcapa} and let  as free parameter the
value the square root of the kinetic-energy density of $q$ at
$\vti$ \cite{jcapa}
\beq\sqrt{\vrho_{{\rm
KI}}}=\vQ(\vti)/\sqrt{2}\simeq\sqrt{\vrho_{q_{\rm
I}}}\simeq\vHi.\eeq
The last equality holds with great accuracy, since we require a
complete domination of kination at early times, as we describe
below.

\subsection{Imposed Requirements}
\label{reqq}

We impose on our quintessential model a number of requirements
which can be described as follows:

\subsubsection{The Constraint of the Initial Domination of Kination.}
As we stress in the introduction, we focus our attention on the
range of parameters that ensure an absolute or at least a relative
initial domination of the $q$-kinetic energy. This requirement can
be quantified as follows:
\beq\Omega^{\rm
I}_q=\Omega_q(\Ti)\gtrsim0.5~~\mbox{with}~~\Omega_q=\rho_q/(\rho_q+\rho_{{\rm
R}}+\rho_{\rm M })\label{domk}\eeq
the quintessential energy-density parameter.

\subsubsection{Nucleosynthesis Constraint.} The
presence of $\rho_q$ has to preserve the successful predictions
of Big Bang NS which starts at about $\vtns=-22.5$
corresponding to $T_{\rm NS}=1~{\rm MeV}$ \cite{oliven}. Taking
into account the most up-to-date analysis of Ref.~\cite{oliven},
we adopt a rather conservative upper bound on $\Omega_q(\vtns)$,
less restrictive than the one of Ref.~\cite{nsb}. In particular,
we require:
\beq\Omega_q^{\rm NS}=\Omega_q(\vtns)\leq0.21~~\mbox{($95\%$
c.l.)} \label{nuc}\eeq
where 0.21 corresponds to additional effective neutrinos species
$\delta N_\nu<1.6$ \cite{oliven}. In the left hand side of
Eq.~(\ref{nuc}), we do not consider extra (potentially large
\cite{giova}) contributions from the energy density of
gravitational waves generated during a possible former transition
from inflation to KD epoch \cite{qinf}. The reason for this
approach is that, inflation could be driven by another field
different from $q$ and therefore, any additional constraint from
that period would be highly model dependent. Nevertheless,
inflation may provide a useful constraint for the parameters of
our model, as we discuss below.

\subsubsection{Inflationary Constraint.\label{resc}} Recent data
\cite{wmap} strongly favors the existence of an inflationary phase
in the early universe. Assuming that this phase also generates the
power spectrum of the curvature scalar $P_{\rm s}$ and tensor
$P_{\rm t}$ perturbations, an upper bound on the inflationary
potential $V_{\rm I}$ and consequently on $H_{\rm I}$ can be
obtained, following the strategy of Ref.~\cite{HIb}. More
specifically, imposing the conservative restriction $r=P_{\rm
t}/P_{\rm s}\lesssim1$, and using the observational normalization
of $P_{\rm s}$ \cite{wmap}, we get
\beq H_{\rm I} \lesssim{\pi\over\sqrt{2}}m_{\rm P}P^{1/2}_{\rm s*}
~~\Rightarrow~~H_{\rm
I}\lesssim2.65\times10^{14}~\GeV~~\Rightarrow~~\vHi\lesssim1.72\times10^{56}
\label{para}\eeq
where $*$ denotes that  $P_{\rm s*}$ is measured at the pivot
scale $k_*=0.002/{\rm Mpc}$. As we can see in Sec.~\ref{ap} the
constraint on $H_{\rm I}$ is sufficient to restrict the parameter
space of our model. Therefore, we are not obliged to impose other
constraint on $\vti$, or equivalently on $\Ti$, as was the case in
Ref.~\cite{jcapa}.

\subsubsection{Coincidence Constraint.} The present value of $\rho_q$,
$\rho^0_q$, must be compatible with the preferred range of
\sEref{cdmba}{b}. This can be achieved by adjusting the value of
$\vVo$. Since, this value does not affect crucially our results
(especially on the \emph{e}-WIMPs abundances), we fix
$\vrho^0_q=\rho^0_q/\rho^0_{\rm c}$ to its central experimental
value, demanding:
\beq \Omega^0_q=\vrho^0_q=0.74.\label{rhoq0}\eeq

\subsubsection{Acceleration Constraint.} A successful
quintessential scenario has to account for the present-day
acceleration of the universe, i.e. \cite{wmap},
\beq-1\leq w_q(0)\leq-0.86~~\mbox{($95\%$ c.l.)}
~~\mbox{with}~~w_q=(\dot q^2/2-V)/(\dot q^2/2+V) \label{wq}\eeq
the barotropic index of the $q$-field. In our case, we are not able
to avoid \cite{german} the eternal acceleration ($w^{\rm
fp}_q>-1/3$ see below), which is disfavored by string theory.

\paragraph{} Let us also comment on our expectations for
the redsift $z_{\rm t}$ of the transition from deceleration to
acceleration, and for the age of the universe, $t_0$. Due to
observational uncertainties in the measurement of these
quantities, we do not impose the data on them as absolute
constraint. We estimate $z_t$ by solving numerically the equation
\beq \mbox{\sf\small q}(\vtau_{\rm t})=0,~~\mbox{with}~~z_{\rm
t}=e^{-\vtauf_{\rm t}}-1~~\mbox{and}~~\mbox{\sf\small q}=-{\ddot R
R/\dot R^2}=-1-{H'/H}\label{zt}\eeq
the deceleration parameter; $t_0$ is also estimated
numerically, as follows:
\beq t_0=\int_{0}^{R_0}
{dR/R}={\left(1/H_0\right)}\int_{\vtauf_{\rm I }}^0
{d\vtau/\vH(\vtau)}\label{t0} \eeq
We obtain $0.78\lesssim z_{\rm t}\lesssim0.84$ and $13.6\gtrsim
t_0/{\rm Gyr}\gtrsim 13.4$ as $w_q(0)$ varies in the range of
\Eref{wq}. In both cases, as $w_q(0)$ approaches $-1$ the results
on $z_{\rm t}$ and $t_0$ approach their values in the context of
the standard power-law cosmological model with CDM and a
cosmological constant ($\Lambda$CDM). The results on $t_0$ are in
agreement with observational data \cite{wmap} (according to which
$t_0=(13.69\pm0.26)~{\rm Gyr}$ at $95\%$ c.l.). On the other hand,
our findings on $z_{\rm t}$ are marginally consistent with the SNe
Ia observations \cite{snae} according to which $z_{\rm
t}=0.46\pm0.26$ at $95\%$ c.l. However, it is probably premature
to say more than that $z_{\rm t}$ is around unity (see, e.g.,
\cref{ztref}).

\subsection{The Quintessential Evolution}
\label{Qev}

The quintessence field $q$ undergoes three phases during its
cosmological evolution \cite{dimopoulos, jcapa}:

\begin{figure}[!t]\vspace*{-.3in}
\begin{center}
\epsfig{file=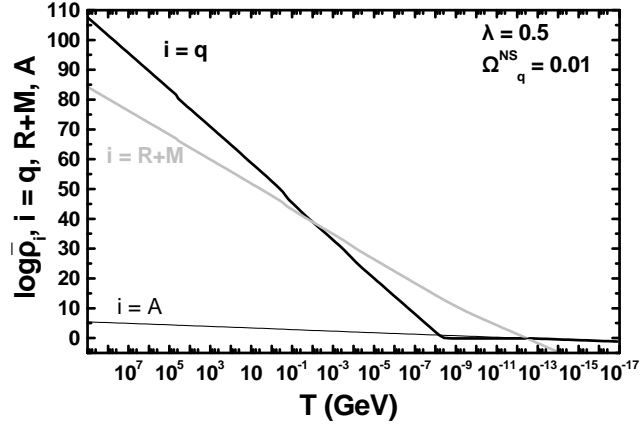,height=3.65in,angle=-90}
\end{center}
\hfill \caption[]{\sl The evolution of the quantities
$\log\vrho_i$ with $i=q$ (black lines), R+M (light gray line) and
A (thin line) as a function of $T$ for $\lambda=0.5$,
$\Ti=10^9~\GeV$ and $\Omqns=0.01~(\vVo=2.04\times 10^{14})$.}
\label{r5}
\end{figure}

\begin{itemize}
\item The kinetic-energy dominated phase during which $\rho_q$ is
essentially given
by $\dot q/2\gg V$; this results to
$w_q\simeq1$ and therefore,
\beq\rho_q=\rho_q^{\rm p}\left({R_{\rm p}\over
R}\right)^6=\rho_q^{\rm p}\left({g_*\over g_*^{\rm
p}}\right)\left({T\over T_{\rm p}}\right)^6 \label{rK} \eeq
(where the last equality was extracted using \Eref{rs}).
The subscript or superscript ``p'' means that the various
quantities are evaluated at a reference point p which can be the
initial epoch (${\rm p=I}$) or the epoch of the transition between
KD and RD era (${\rm p=KR}$) or the eve of NS (${\rm p=NS}$).
Applying \Eref{rK} for ${\rm p=NS}$, we derive the "transition"
temperature $\Tkr$ from KD to RD era as follows:
\beq\rho_q(\Tkr)=\rho_{\rm
R}(\Tkr)~\Rightarrow~\Tkr=\Tns\left({g^{\rm NS}_{*}\over g^{\rm
KR}_{*}}\right)^{1/2}\left({1-\Omega^{\rm NS}_q \over\Omega^{\rm
NS}_q}\right)^{1/2}\cdot\label{Tkr} \eeq
For $T>\Tkr$, the universe undergoes a KD epoch where $H$
is adequately approximated by
\beq H\simeq{1\over \sqrt{3}m_{\rm P}}\rho^{1/2}_{\rm
R}\left(1+{\Omega^{\rm p}_q\over1-\Omega^{\rm p}_q}\left({g_*\over
g_*^{\rm p}}\right)\left({T\over T_{\rm
p}}\right)^2\right)^{1/2}\cdot \label{Hkin} \eeq
Applying this formula for ${\rm p=NS}$ for any given $\Ti$, we get
$H_{\rm I}$ as a function of $\Omqns$. Therefore, we can use
$\Omqns$ as a free parameter,  instead of $H_{\rm I}$.
\item The frozen-field dominated phase, where the
universe becomes RD and $\rho_q$ is dominated initially by $\dot
q/2$ and subsequently by $V$.

\item The attractor dominated phase, where $\rho_q\simeq V$
dominates the evolution of the universe, and reaches the late-time
attractor energy density:
\beq\vrho_{{\rm A}}\simeq\vrho_q^0
\left({R/R_0}\right)^{-3(1+w^{\rm fp}_q)}~~\mbox{with}~~w^{\rm
fp}_q=\lambda^2/3-1~~(\mbox{for} ~~\lambda<\sqrt{3})
\label{attr}\eeq
the fixed point value of $w_q$. Today we obtain a transition from
the frozen-field dominated phase to the attractor dominated phase
\cite{jcapa, brazil}. Although this is not a satisfactory solution
to the coincidence problem, the observational data can be
reproduced for a reasonable set of initial conditions, as we show
below.
\end{itemize}

\begin{figure}[!t]\vspace*{-.3in}
\begin{center}
\epsfig{file=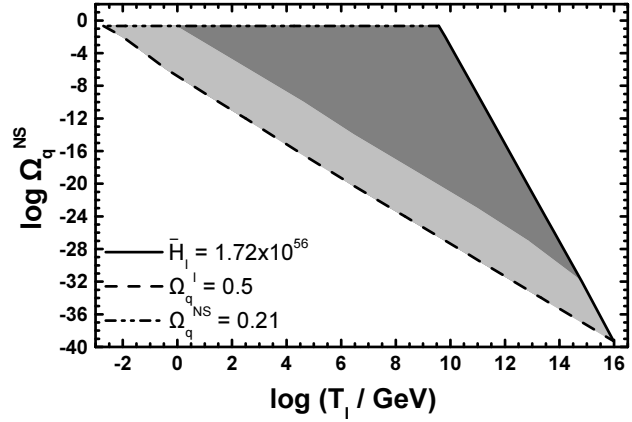,height=3.65in,angle=-90}
\end{center}
\hfill \caption[]{\sl Allowed (gray and lightly grey shaded)
region in the $\log\Ti-\log\Omqns$ plane by Eqs.~(\ref{domk}),
(\ref{nuc}) and (\ref{para}). The conventions adopted for the
various lines are also shown.} \label{OmT}
\end{figure}

The cosmological evolution described above is illustrated in
\Fref{r5} where we plot $\log\vrho_i$ versus $T$ for
$\Ti=10^9~\GeV$ ($\vti=-51.2$), $\Omqns=0.01$
($\vHi=7.07\times10^{53}$) and $\lambda=0.5~(\vVo=2.04\times
10^{14})$. For $i=q$ (bold black line), we show $\log\vrho_q$,
computed by inserting in Eq.~(\ref{vrhoq}) the numerical solution
of Eq.~(\ref{vH}). For $i={\rm A}$ (thin black lines), we show
$\log\vrho_{{\rm A}}$ derived from Eq.~(\ref{attr}). For $i={\rm
R+M}$ (light gray line), we show $\log\vrho_{{\rm R+M}}$, which is
the logarithm of the sum of the contributions given by
Eq.~(\ref{rhos}). For $T>\Tkr=0.0098~\GeV$ we have a KD era. We
also obtain $w_q(0)=0.957$ whereas $w_q^{\rm fp}=0.916$, $z_{\rm
t}=0.8$ and $t_0=13.5~{\rm Gyr}$.

\subsection{The Allowed Parameter Space}
\label{ap}

The free parameter space of our quintessential model is defined
by the following quantities:
$$\lambda,\ \vti~~\mbox{and}~~\vHi ~~\mbox{or equivalently,}~~
\lambda,\ \Ti~~\mbox{and}~~\Omqns.$$
Agreement with Eq.~(\ref{wq}) entails $0<\lambda\lesssim0.9$
(compare with \cref{german, jcapa}, where less restrictive
upper bounds on $w_q(0)$ have been imposed)). This range is
independent on $\vti$ and $\vHi$ (or $\Ti$ and $\Omqns$) as is
shown in \cref{jcapa}. Given this fact, we focus on the last two
free parameters of our model.

The allowed area in the $\log\Ti-\log\Omqns$ plane (which obviously is
$\lambda$-independent), is indicated in Fig.~\ref{OmT}.
In the shaded areas Eqs.~(\ref{domk})
(\ref{nuc}) and (\ref{para}) are fulfilled; in particular, in the
dark [light] shaded area, $\Omega_q^{\rm I}=1$
[$0.5\lesssim\Omega_q^{\rm I}\lesssim1$]. The left [right]
boundary of the allowed regions (indicated by a dashed [solid]
line) is derived from \Eref{domk}~[\Eref{para}] whereas the upper
boundary (indicated by a double dot-dashed line) comes from
\Eref{nuc}. We observe that the dashed and solid lines are almost
straight. This can be understood by deriving the analytic
relation between $\Ti$ and $\Omqns$ for fixed $H_{\rm I}$ or
$\Omega_q^{\rm I}<1$, namely:
\beq \label{Tomq} \Ti=\Tns\sqrt{{g_*^{\rm NS}\over g_*^{\rm I}}
{1-\Omqns\over \Omqns}\left({\vHi^2\over\vrho_{\rm R}^{\rm
I}}-1\right)}~~\mbox{and}~~\Ti=\Tns\sqrt{{g_*^{\rm NS}\over
g_*^{\rm I}} {1-\Omqns\over \Omqns}{\Omega_q^{\rm I}\over
1-\Omega_q^{\rm I}}}\cdot\eeq
The first [second] equation can be extracted by calculating
$H_{\rm I}$ through \Eref{Hkin} with ${\rm p=NS}$ [${\rm p=I}$ and
${\rm p=NS}$ and eliminating $H_{\rm I}$] and solving the
resulting w.r.t $\Ti$. Taking the logarithms of the two sides of
\Eref{Tomq} and given that $\Omega_q^{\rm NS}$ is quite small in
the largest part of the available parameter space, we can convince
ourselves that the dashed and solid curves in the
$\log\Ti-\log\Omqns$ plot have to be almost straight.
Consequently, for a reasonable set of parameters ($\lambda, \Ti,
\Omqns$), the exponential quintessential model can become
consistent with the observational data, in agreement with
\cite{jcapa, brazil, german}.

\section{Thermal Abundance of SUSY \emph{e}-WIMPs}\label{sec:boltz}
\setcounter{equation}{0}

We assume that any relic abundance of an \emph{e}-WIMP, $\chia$,
due to its decoupling from the thermal bath, is diluted after
inflation at an energy scale below $10^{10}~\GeV$. We compute its
abundance produced through thermal scatterings and decays during
the KD epoch and compare this result with the one obtained in the
SC. The relevant equations are presented in Sec.~\ref{Beq}. In
Sec.~\ref{Neqs} we describe our numerical evaluation and in
Sec.~\ref{Seqs} we derive useful approximate expressions. Their
results are compared with the numerical ones in Sec.~\ref{numan}.

\subsection{The Boltzmann Equation}
\label{Beq}

The number density $n_{\chia}$ of $\chia$ particles satisfies the
Boltzmann equation, which can be collectively written in the
following form \cite{moroi1, axino}:
\beq \dot n_\chia+3Hn_\chia={1\over2}\sum_{ij}\sigv{ij}
\nequ_i\nequ_j+\sum_i\Gma{i} \nequ_i. \label{nx}\eeq
Here, $H$ is given by Eq.~(\ref{rhoqi}) and $n_i^{\rm eq}$ is the
equilibrium number density of the particle $i$, which can be
adequately calculated in both the relativistic ($m_i\ll T$) and
non-relativistic ($m_i\gg T$) regime \cite{wolfram}:
\begin{equation} \label{neq}
\nequ_i=\frac{g_i}{2\pi^2} m^2_i T K_2(m_i/T)
\end{equation}
with $K_n$, the modified Bessel function of the second kind of
order $n$ and $m_i$ [$g_i$] the mass [the number of degrees of
freedom] of the particle $i$.

The quantity $\sigv{ij}$ is the thermal-averaged production rate
of $\chia$ from scatterings in the thermal plasma, the indices $i$
and $j$ run over all particles involved in the initial states of
these processes, namely gluons ($g$), gluinos ($\gl$), quarks
($q$) and squarks ($\sq$), and the factor $1/2$ is introduced
\cite{moroi1} in order to avoid double counting of the production
processes. Formalistically speaking, the term
$\sigv{ij}\nequ_i\nequ_j/2$ corresponds to the so-called in
Refs.~\cite{Bolz, steffen, steffenaxino} collision term $C_\chia
n^{{\rm eq}2}$ and therefore, we can make the replacement:
\bea \label{sig}
{1\over2}\sum_{ij}\sigv{ij}\nequ_i\nequ_j=C_{\chia} n^{{\rm
eq}2}~~\mbox{with}~~C_{\chia}=\left\{\matrix{C_\chia^{\rm HT}&
\mbox{in the HTR,} \cr C_\chia^{\rm LT}&\mbox{in the LTR,} \hfill
\cr} \right. \label{neqb}\eea
and $\nequ=\zeta(3)T^3/\pi^2$, the equilibrium number density of
the bosonic relativistic species ($\zeta(3)\simeq1.2$ is the
Riemann zeta function of 3).

We proceed with the formulas for $C_\chia$'s in the low (LTR) and
high (HTR) $T$ regime. In the relativistic regime ($T\gg m_i$),
$C_\chia$ has been recently recalculated \cite{Bolz, steffen,
steffenaxino} in a consistent gauge-invariant treatment, using
\emph{Hard Thermal Loop Approximation} (HTLA) technics. The result
is
\beq \hspace{-1cm}C_\chia^{\rm HT} =\left\{\matrix{
\hspace{-2.5cm}\begin{minipage}[h]{10cm}\vspace*{-.3cm}
\bea\nonumber {3\pi\over16\zeta(3)m^2_{\rm P}}\sum_{\alpha=1}^{3}
\left(1+{M_\alpha^2\over3 m_{\chia}^2}\right)c_\alpha g_\alpha^2
        \ln\left({k_{\alpha}\over g_\alpha}\right)
\eea\end{minipage} \hfill \mbox{for}  & \chia=\Gr, \hfill \cr
\hspace{-2.5cm}\begin{minipage}[h]{10cm}
\vspace*{-.3cm}\bea\nonumber {108\pi g_a^2 g_3^2\over \zeta(3)}
\ln\left({1.108\over g_3}\right)~~
\mbox{with}~~g_a={g_3^2\over32\pi^2f_a}\eea\end{minipage} \hfill
\mbox{for} & \chia=\tilde a. \hfill \cr}
\right. \label{sig1}\eeq
Here, $g_\alpha$ and $M_\alpha$ are the gauge coupling constants
and gaugino masses respectively, associated with the gauge groups
$U(1)_{\rm Y}$, $SU(2)_{\rm L}$ and $SU(3)_{\rm C}$,
$(k_\alpha)=(1.634,1.312,1.271)$ and $(c_\alpha)=(33/5,27,72)$.
Note that, contrary to the notation of \cref{steffen}, we use the
\emph{Grand Unified Theory} (GUT) inspired normalization of the
hypercharge coupling constant $g_1$.

In our study, we calculate $g_\alpha$  as a function of
the temperature, solving the relevant one loop renormalization group
equations with the MSSM particle content. In particular, we have
\beq g_\alpha(T)=g_{\rm GUT}\left({1\over g^2_{\rm
GUT}}-{b_\alpha\over8\pi^2} \ln {T\over M_{\rm
GUT}}\right)^{-1/2}\label{gs}\eeq
where $(b_\alpha)=(33/5,1,-3)$ and $g_{\rm GUT}\simeq1/24$ is the
value of $g$'s at the GUT scale $M_{\rm
GUT}\simeq2\times10^{16}~\GeV$. By adjusting $g_{\rm GUT}$ and
$M_{\rm GUT}$, we obtain the experimentally acceptable values
of $g_\alpha$'s at $M_Z$. Below $M_Z$, we use the $g_3$ running
indicated in \cref{qcdg3}.

Throughout our analysis we impose universal initial conditions for
the gaugino masses i.e., we assume that
\beq M_\alpha(M_{\rm
GUT})=M_{1/2},~~\mbox{with}~~\alpha=1,2,3.\label{Mg}\eeq
The running of the gaugino masses $M_\alpha$ can be easily
evaluated at one loop by solving the relevant renormalization
group equations, which admit an exact solution:
\beq M_\alpha(T)=M_{1/2}{\left(g_\alpha \over g_{\rm
GUT}\right)^2}=M_{1/2}\left(1-{b_\alpha\over8\pi^2}\ln {T\over
M_{\rm GUT} }\right)^{-1}\label{M12}\eeq
Under the assumption of \Eref{Mg}, the lightest neutralino turns
out to be a Bino ($\tilde B$), with mass $m_{\tilde
B}\simeq0.41\,M_{1/2}$.

From \Eref{sig1} one easily deduces that $C^{\rm HT}_\chia$
becomes negative for large enough values of $g_3$. Indeed, from
\Eref{gs} we conclude that $g_3$ increases as $T$ decreases. In
particular, $C^{\rm HT}_{\ax}>0$ requires $g_3 \lesssim 1.108$ or
$T>\Tc=10^4~\GeV$. \Eref{sig1}, therefore, can be applied
self-consistently for $\Ti>\Tc$ and $\Tkr>\Tc$. Towards lower
values of $T$, the finite masses $m_i$, which have been neglected
in deriving \Eref{sig1}, and $\chia$-production from decays, start
playing an important role. The correct inclusion of these effects
is crucial in the case of $\ax$, since $\Omax$ takes
cosmologically interesting values for $\Ti<\Tc$ and/or $\Tkr<\Tc$,
too. In Appendix A, we present a novel calculation of the term
$\sigv{ij}\nequ_i \nequ_j$ in the case of $\ax$ and in the
non-relativistic regime ($T\ll m_i$). Our final result can be cast
in a form similar to this of \Eref{sig}, with:
\begin{equation}
C^{\rm LT}_{\ax}= {1\over 16 T^5\zeta(3)^2}
\sum_{ij}\int_{s_0}^\infty ds\,w_{ij}(s)K_1\left({\sqrt{s}\over
T}\right)p_{\rm i}(m_i,m_j)\;, \label{sig2}
\end{equation}
where $s_0=(m_i+m_j)^2$ and the symbols $w_{ij}$ and $p_{\rm i}$
are defined in Eqs.~(\ref{wij}) and (\ref{ps1}). In particular,
$w_{ij}$ is related to the squared amplitudes of the various
processes that contribute to $\sigv{ij}$, and are listed in Table
1.

Finally, $\Gma{i}$ is the thermal-averaged rate of
$\chia$-production from decays in the thermal plasma, which are
related to the corresponding decay widths $\Gm{i}$ through the
formula \cite{wolfram}:
\beq \label{gma} \Gma{i}={K_1(m_i/T)\over
K_2(m_i/T)}\Gm{i}~\Rightarrow~\Gma{i}\nequ_i={g_i\over2\pi^2}m_i^2\,\Gm{i}
T\,K_1(m_i/T)\eeq
These contributions have been recently calculated in the case
of $\Gr$ in \cref{strumia}; however, we do not include them
in our calculation since their impact is roughly
a factor of two. Such a minor change does not alter
our results in any essential way. On the other hand, we do include
these contributions in the case of $\ax$ with $i=\gl,~\sq$ and
$\tilde B$, applying the following formulas \cite{small}:
\beq \label{gms}\Gm{i} =\left\{\matrix{
\hspace{-2.5cm}\begin{minipage}[h]{7.6cm}
\vspace*{-.3cm}\bea\nonumber
4{g^2_ag_3^2\over\pi}m^3_{\gl}\left(1-{m^2_{\ax}\over
m^2_{\gl}}\right)^3 \eea\end{minipage}\hfill \mbox{for}& i=\gl,
\hfill \cr
\hspace{-2.5cm}\begin{minipage}[h]{7.6cm}
\vspace*{-.3cm}\bea\nonumber
{3\over2}{g^4_ag_3^4\over\pi^5}m_{\sq}\left(m_{\gl}\ln{f_{a}\over
m_{\gl}}\right)^2 \eea\end{minipage}\hfill  \mbox{for} & i=\sq,
\hfill \cr
\hspace{-2.5cm}\begin{minipage}[h]{7.6cm}
\vspace*{-.3cm}\bea\nonumber {9g_1^4\over50\times
32^2\pi^5f_a^2}m^3_{\tilde B}\left(1-{m^2_{\ax}\over m^2_{\tilde
B}}\right)^3 \eea\end{minipage}\hfill \mbox{for}& i=\tilde B.
\hfill \cr}
\right. \eeq
In the above equations, $g_3$ is calculated at a scale equal to $m_i$ if
$\Ti>m_i$ or $\Ti$ if $\Ti<m_i$. Obviously, $\Gamma_{\gl}$ and
$\Gamma_{\sq}$ are much more efficient than $\Gamma_{\tilde B}$,
due to the presence of $g_3$. As we verify numerically, the
contribution of these terms to the resulting $\Omx$ becomes
important \cite{small, axino, ewimps} for $\Ti<T_{\rm SUSY}$ or
$\Tkr<\Ts$, where $T_{\rm SUSY}\simeq m_{\sq}$.

\subsection{Numerical Solution}
\label{Neqs}

In order to find a precise numerical solution to our problem, we
have to solve \Eref{nx} together with \Eref{qeq}. To this end, and
following the strategy of Sec.~\ref{Beqs}, we introduce the
dimensionless quantities:
\beq \bar n_\chia=n_\chia/\left(\rho^0_{\rm
c}\right)^{3/4}~~\mbox{and}~~\bar n^{\rm eq}=n^{\rm
eq}/\left(\rho^0_{\rm c}\right)^{3/4}.\eeq
In terms of these quantities (and substituting Eqs.~(\ref{sig}) and
(\ref{gma}) into Eq.~(\ref{nx})), this takes the following master form,
for numerical manipulations:
\beq \vH\vn^\prime_\chia+3\vH \vn_{\chia}-\bar C_\chia\vn^{\rm
eq2}-\sum_i{g_i\over2\pi^2}\bar\Gm{i}\bar m_i^2 \bar
T\,K_1(m_i/T)=0,\label{rx}\eeq
where $\vH$ is given in Eq.~(\ref{vH}) and the following
quantities have been defined:
\beq \bar C_\chia=C_\chia\sqrt{3}m_{\rm P}\left(\rho^0_{\rm
c}\right)^{1/4},~\bar m_i={m_i \over\left(\rho^0_{\rm
c}\right)^{1/4}},~\bar T={T \over\left(\rho^0_{\rm
c}\right)^{1/4}}~~\mbox{and}~~\bar\Gm{i}={\Gm{i}\over H_0}.\eeq
Eq.~(\ref{rx}) can be solved numerically with the initial
condition $\vn_\chia(\vti)\simeq0$, where $\vti$ corresponds to
the initial temperature $\Ti$. The integration of \Eref{rx} runs
from $\Ti$ down to $\Tns$ (an integration to $0$ also gives the
same result). We find convenient to single out two cases:

\begin{itemize}

\item In the HTR ($\Ti\gg \Tc$ and $\Tkr\gg \Tc$), we integrate
\Eref{rx} with $C_\chia=C^{\rm HT}_\chia$ from $\Ti$ to $\Tns$ (in
the integration for $T<\Tc$ we take $C^{\rm HT}_\chia$ frozen at
its value at $T\simeq\Tc$). As it turns out, contributions from
Eqs.~(\ref{sig2}) and (\ref{gms}) are negligible.

\item In the LTR ($\Ti\ll\Tc$ or $\Tkr\ll\Tc$), if (i) $\Ti>\Ts$,
we integrate \Eref{rx} successively from $\Ti$ to $\Ts$ with
$C_{\ax}=C^{\rm HT}_{\ax}$ and then from $\Ts$ to $\Tns$ with
$C_{\ax}=C^{\rm LT}_{\ax}$ whereas if (ii) $\Ti<\Ts$ we integrate
\Eref{rx} from $\Ti$ to $\Tns$ with $C_{\ax}=C^{\rm LT}_{\ax}$. It
turns out that the contributions from Eq.~(\ref{gms}) are
important only for $T<\Ts$. Note that for $\Ts<T<\Tc$ (where the
SUSY particles $i$ with masses $m_i\sim\Ts$ are relativistic)
there is no accurate result for $\sigv{ij}$, since neither HTLA is
valid ($g_3(T)>1.1$) nor our computation in Appendix A is
applicable (we consider non-relativistic particles $i$ and $j$).
However, we believe that the above procedure gives a result which
is sufficiently accurate for our proposes, since small variation
of $\Ts$ (by 10$\%$) leaves the result practically unaltered.

\end{itemize}

The $\chia$ yield, $Y_\chia=n_{\chia}/s$, and the relic density of
$\chia$, $\Omega_{\chia}h^2=\rho_{\chia}^0 h^2/\rho^0_{\rm c}$
(with $\rho_\chia=m_\chia n_\chia$), can be easily found, via the
relations \cite{edjo}:
\beq \label{omega} \mbox{\sf\small (a)}
~Y^0_\chia=3.55\times10^{-27}\ \bar
n_{\chia}(\Tns)~~\mbox{and}~~\mbox{\sf\small (b)}~ \Omx= 2.748
\times 10^8\ Y^0_\chia\ m_{\chia}/\GeV,\eeq
where $\left(\rho^0_{\rm
c}\right)^{3/4}/s(\Tns)=3.55\times10^{-27}$ and $s_0 /\rho_{\rm
c}^0= 2.748 \times 10^8/\GeV$.

\subsection{Semi-Analytical Approach}\label{Seqs}

The crucial quantity $Y^0_\chia$ for the computation of $\Omx$ via
\Eref{omega} can be also derived semi-analytically by employing a
number of simplifications. We first re-express Eq.~(\ref{nx}) in
terms of the variable $Y_\chia = n_\chia/s$, in order to absorb
the dilution term. Indeed \cite{edjo}
\beq\label{eq1} \dot n_{\chia}+3H\, n_{\chia}=\dot Y_\chia
s=Y_\chia^{\prime}\dot T s.\eeq
where prime in this section means derivative w.r.t $T$. Employing
\Eref{rs}, we obtain:
\begin{equation}
\dot s= -3Hs \Rightarrow \dot T=-3Hs/s^\prime, \label{eq2}
\end{equation}
where $H$ can be fairly approximated by applying \Eref{Hkin} with
p=KR:
\beq \label{eq3} H\simeq{\rho^{1/2}_{\rm R}\over\sqrt{3}m_{\rm
P}}(1+r_q)^{1/2},~~\mbox{with}~~r_q={g_*\over g^{\rm
KR}_*}\left({T\over \Tkr}\right)^2\eeq
The presence of $r_q>0$ clearly indicates the deviation from
the SC, where $r_q=0$ and $H=H_{\rm SC}=\sqrt{\rho_{\rm
R}}/\sqrt{3}m_{\rm P}$. Inserting Eqs.~(\ref{eq2}) and(\ref{eq3})
into Eq.~(\ref{eq1}) and substituting in Eq.~(\ref{nx}), this can
be rewritten in terms of the new variables as:
\beq \label{eq4} Y_\chia=Y_\sigma+Y_\Gamma\eeq
where $Y_\sigma$ and $Y_\Gamma$ satisfy the following differential
equations:
\beq \label{Ys} Y_\sigma^{\prime}
=-y_\sigma/\sqrt{1+r_q}~~\mbox{and}~~Y_\Gamma^{\prime} =
-\sum_iy_{\Gamma_i}K_1(m_i/T)/T^{5}\sqrt{1+r_q}\, .\eeq
In the above, we have defined the following quantities:
\bea \label{ys} && y_\sigma=y_{\chia*}Y^{\rm
eq2}C_\chia~~\mbox{and}~~y_{\Gamma_i}=
{\pi^2\over2\zeta(3)^2}Y^{\rm eq2}y_{\chia*}g_i\Gamma_im_i^2
~~\mbox{with}~~Y^{\rm eq}={n^{\rm eq}\over s},\\ &&
y_{\chia*}={s'\over3H_{\rm SC}}=\sqrt{\frac{8}{45}}\pi m_{\rm P
}g_{\chia*}^{1/2}~~\mbox{and \cite{edjo}}~~g_{\chia*}^{1/2} =
\frac{g_{s*}}{\sqrt{g_{\rho*}}} \left( 1+\frac{T
g^\prime_{s*}}{3g_{s*}} \right)\cdot \eea
The expression of $H$ as a function of $\Tkr$ in \Eref{eq3}
enables us to divide the integration of Eqs.~(\ref{Ys}) to two
separated domains, since $1/\sqrt{1+r_q} $ can be approximated as
follows (compare with Refs.~\cite{seto, panot}):
\beq\label{rqap} 1/\sqrt{1+r_q} \simeq\left\{\matrix{
\sqrt{{g^{\rm KR}_*/ g_*}}\left(\Tkr/T\right)\hfill &\mbox{for}~~
T\gg\Tkr,\cr
1\hfill & \mbox{for}~~T\ll\Tkr.\hfill \cr}
\right. \eeq
Following this simplification, \Eref{Ys} can be resolved trivially.
Let us present our results for the HTR and the LTR separately.

\subsubsection{The HTR.} In this case ($\Ti\gg \Ts$ and $\Tkr\gg \Ts$),
we find that $Y_\chia^0\simeq Y_\sigma^0$ where
\beq \label{Ysol} Y^0_\sigma=\int_{\Tkr}^{\Ti} dT\sqrt{{g^{\rm
KR}_*\over g_*}} {\Tkr\over T}y^{\rm HT}_\sigma +\int^{\Tkr}_0 dT
y^{\rm HT}_\sigma \eeq
and $y^{\rm HT}_\sigma=y_\sigma(C^{\rm HT}_\chia)$. Fixing $C^{\rm
HT}_\chia$ and $g_*$ at their initial values in each domain of
integration, an analytic formula can be derived, within an $10\%$
accuracy:
\beq \label{YhT} Y_\sigma^0\simeq\left\{\matrix{ y^{\rm HT}_\sigma
\Ti\hfill & \mbox{in the SC}, \hfill \cr y^{\rm HT}_\sigma
\sqrt{{g^{\rm KR}_*/ g^{\rm I}_*}}\Tkr\ln\left(T_{\rm I}/T_{\rm
KR}\right)+ y^{\rm HT}_\sigma T_{\rm KR} \hfill &\mbox{in the
QKS}. \hfill \cr}\right. \eeq
From the above expressions, we can easily deduce that $Y_\chia^0$
decreases in the QKS w.r.t its value in the SC. This is to be
expected, since in the SC it is proportional to $\Ti$ whereas, in
the QKS, it becomes proportional to $\Tkr$ (lower than $\Ti$). In
the latter case, the first term (which is usually ignored in
similar estimates \cite{seto, panot}) of the \emph{right hand
side} (r.h.s.) of \Eref{YhT}, gives an equally important
contribution.

\subsubsection{The LTR.} In this case ($\Ti\ll \Tc$ or $\Tkr\ll
\Tc$), we find cosmologically interesting solutions only in the
case of $\ax$. We focus on the most intriguing possibility, in
which $\Ti\gg\Ts$ but $\Tkr\ll\Ts$. In this case, $\Omqns$ takes
naturally a value close to its upper bound in \Eref{nuc}, as can be
verified via \Eref{Tkr}. We find
\bea \label{Ya} Y^0_{\ax}&= &Y^{(1)}_{\sigma}+ Y^{(2)}_{\sigma}+
Y^0_{\Gamma}~~\mbox{where}\\
\label{Ya1} Y^{(1)}_{\sigma}&\simeq&\int^{\Ti}_{\Ts}
dT\sqrt{{g^{\rm KR}_*\over g_*}}{\Tkr\over T}y^{\rm
HT}_\sigma,\\
\label{Ya2} Y^{(2)}_{\sigma}&\simeq&\int^{\Ts}_{\Tkr} dT
\sqrt{{g^{\rm KR}_*\over g_*}}{\Tkr\over T}y^{\rm LT}_\sigma
+\int^{\Tkr}_0 dT\,
y^{\rm LT}_\sigma, \\
Y^0_{\Gamma}&\simeq &\sum_i\left(\int^{\Ts}_{\Tkr} dT
\sqrt{{g^{\rm KR}_*\over g_*}}{\Tkr\over T^6}+\int^{\Tkr}_0
{dT\over T^{5}} \right)y_{\Gamma_i} K_1\left({m_i\over
T}\right),\label{Ya3}\eea
%
with $y^{\rm LT}_\sigma=y_\sigma(C^{\rm LT}_\chia)$. In the SC
with $\Ti<\Ts$, $Y^0_{\ax}$ can be derived by summing the second
terms of the r.h.s. of Eqs.~(\ref{Ya2}) and (\ref{Ya3}) and
setting $\Tkr=\Ti$. Unfortunately, finding a general analytical
expression for Eqs.~(\ref{Ya1}) -- (\ref{Ya3}) is not
straightforward, mainly because the derivation of $C^{\rm
LT}_{\ax}$ from \Eref{sig2} requires a double integration of
several lengthy squared amplitudes (see Appendix A). However, for
the benchmark values of $m_i$ used in our analysis:
\beq \label{mi}
m_{\sq}=1~\TeV,~m_{\gl}=1.5~\TeV~~\mbox{and}~~m_{\tilde
B}=0.3~\TeV \eeq
we can write simple empirical relations which reproduce rather
accurately our numerical results. We distinguish the following cases:

\begin{itemize}

\item In the SC, the main contribution to $Y_{\ax}^0$ arises from
the second term of the r.h.s. of \Eref{Ya2} [\Eref{Ya3}] with
$\Tkr=\Ti$ for $\Ti>0.3~\TeV$ [$\Ti<0.3~\TeV$]. Using fitting
technics, we get a relation with a $15\%$ accuracy:
\beq\label{emp0}\Omax=A\, m_{\ax}\,\left(1 +C\,\Ti
\right)\,e^{-B/\Ti}/f^2_a~~\mbox{with}~~A=1.44\times
10^{24}~\GeV,\eeq
$B=745.472~\GeV$ and $C=0.001/\GeV$. The discrepancy can be
attributed to the logarithmic factor involved in $\Gm{\sq}$ -- see
\Eref{gms}. This factor disturbs the dependence of $\Omax$ on
$f_a$ as written in \Eref{emp0}. The origin of the exponential
factor is the non-relativistic expansion \cite{wolfram} of
$K_1(\sqrt{s}/T)$ [$K_1(m_i/T)$] involved in the definition of
$\sigv{ij}$ [$\Gma{i}$] -- see \Eref{sig2} [\Eref{gma}].

\item In the QKS, the main contribution to $Y_{\ax}^0$ comes from
the first term of the r.h.s. of \Eref{Ya2} and therefore,
we expect that $\Omax$ is independent of $\Ti$.
Our results can be reproduced
from the following relation which holds with an excellent
accuracy:
\beq\label{emp}\Omax={D\over f^2_a}\,
{m_{\ax}\over\sqrt{\Omqns}}={D\over
f^2_a}{\Tkr\over\Tns}{\sqrt{g_*^{\rm KR}\over g_*^{\rm
NS}}}\,m_{\ax}~~\mbox{with}~~D=9.26\times10^{17}~\GeV
\eeq
where we used \Eref{Tkr} in the last equality. We observe that
$\Omax\propto1/\sqrt{\Omqns}$ or $\Omax\propto\Tkr$.

\end{itemize}
We conclude, therefore, that in both the QKS and the SC our calculation of
$C^{\rm LT}_{\ax}$ is crucial in order to achieve a reliable
result for $\Omax$.

\subsection{Numerical Versus Semi-Analytical Results}
\label{numan}

\begin{figure}[!t]\vspace*{-.28in}
\hspace*{-.25in}
\begin{minipage}{8in}
\epsfig{file=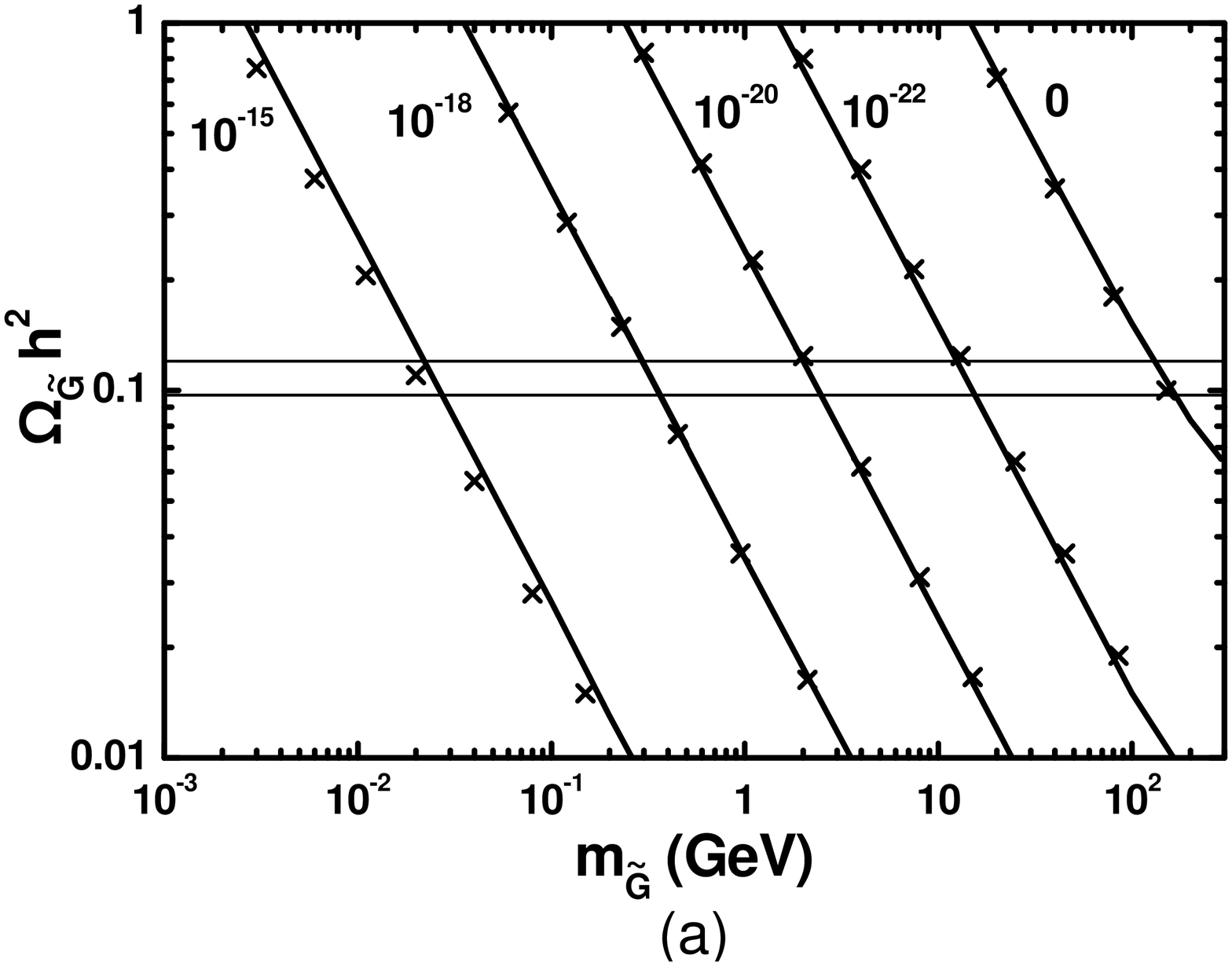,height=3.65in,angle=-90}
\hspace*{-1.37 cm}
\epsfig{file=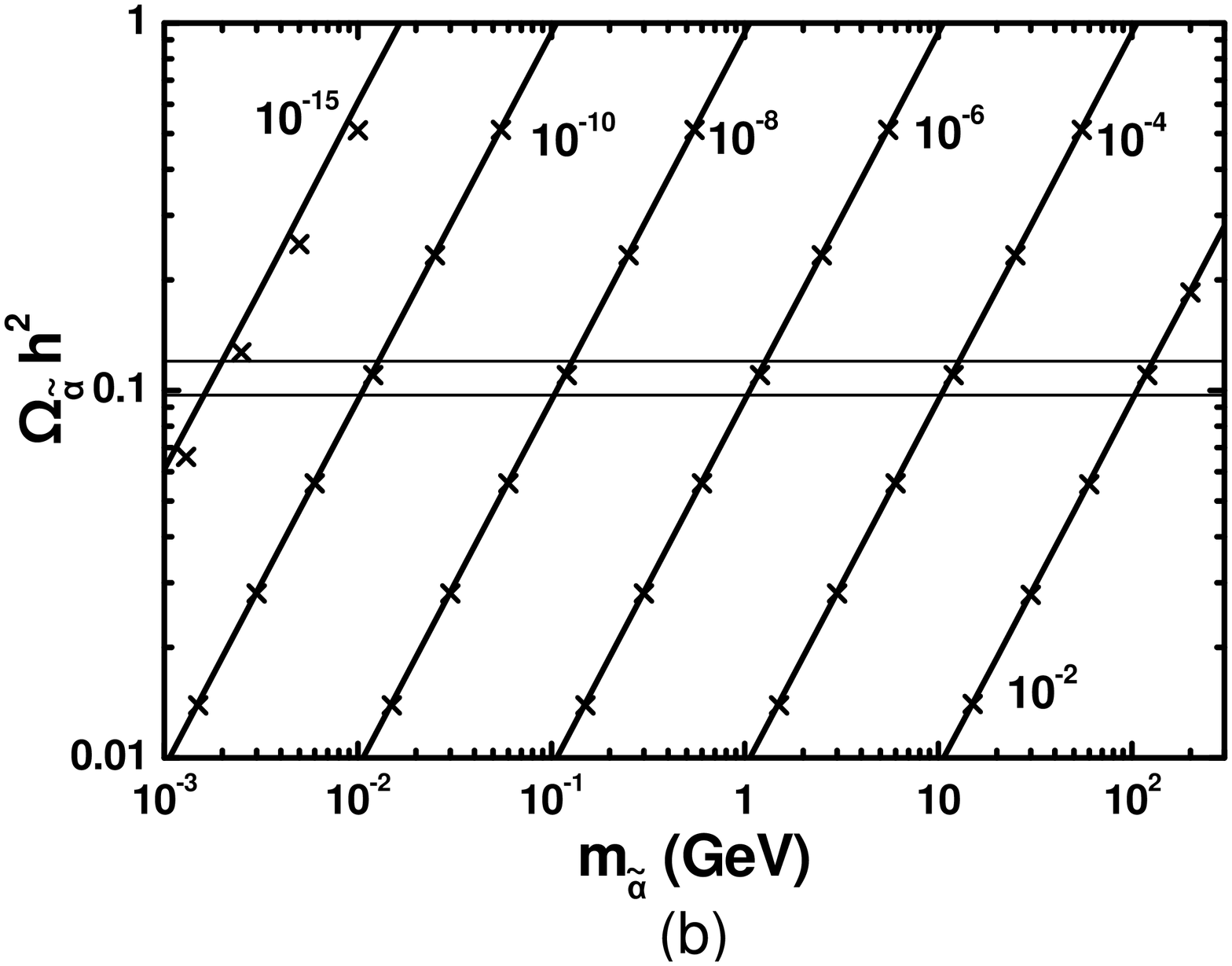,height=3.65in,angle=-90} \hfill
\end{minipage}
\hfill \caption[]{\sl $\Omega_{\chia}h^2$ as a function of
$m_\chia$ ($\chia=\Gr$ [$\chia=\ax$]) for various $\Omega_q^{\rm
NS}$'s, indicated on the curves, $\Ti=10^9~\GeV$ and
$M_{1/2}=0.7~\TeV$ [$f_a=10^{11}~\GeV$] ({\sf\ssz a} [{\sf\ssz
b}]). For $\Omega_q^{\rm NS}>10^{-15}$, we take in our computation
the values of $m_i$ indicated in \Eref{mi}. The solid lines
[crosses] are obtained by our numerical code [semi-analytical
expressions]. The CDM bounds of Eq.~(\ref{cdmb}) are also,
depicted by the two thin lines.}\label{om}
\end{figure}

The validity of our semi-analytical approach can be tested by
comparing its results for $\Omx$ with those obtained by the
numerical solution of \Eref{rx}. In addition, useful conclusions
can be inferred for the behavior of $\Omx$ as a function of
$m_\chia$ and $\Omega_q^{\rm NS}$. Our results are presented in
\Fref{om}. The lines are drawn applying our numerical code,
whereas crosses are obtained by employing the formulas of
Sec.~\ref{Seqs}. In Figs.~\ref{om}-{\sf\small (a)}
[\ref{om}-{\sf\small (b)}], we display $\Omgr$ [$\Omax$] versus
$m_{\Gr}$ [$m_{\ax}$] for $M_{1/2}=0.7~\TeV$ [$f_a=10^{11}~\GeV$]
and various $\Omqns$'s indicated on the curves. We take
$\Ti=10^9~\GeV$, a value frequently met in the well-motivated
models of SUSY hybrid inflation \cite{susyhybrid}.

As we can verify via \Eref{Tkr}, the results for the HTR are
applicable for any $\Omqns$ in \sFref{om}{a} and for
$\Omqns=10^{-15}$ in \sFref{om}{b}, whereas for the residual
$\Omqns$ in \sFref{om}{b} the results for LTH hold; in particular,
the crosses are obtained from \Eref{Ysol} [\Eref{Ya}] in the HTR
[LTR]. In both regimes, $\Omega_{\chia}h^2$ decreases as
$\Omega_q^{\rm NS}$ increases, as we expect by combining
Eqs.~(\ref{Ysol}) and (\ref{Tkr}) for the HTR, and from \Eref{emp}
for the LTR. On the other hand, $\Omgr\propto 1/m_{\Gr}$, while
$\Omax\propto m_{\ax}$. This can be understood from the fact that
$\Omx\propto m_{\chia}Y_\chia^0$ and $Y_\chia^0\propto C_\chia$ or
$Y_\chia^0\propto \Gm{i}$. However $C^{\rm HT}_{\Gr}\propto
1/m^2_{\Gr}$ (for $M_\alpha>m_{\Gr}$) whereas $C^{\rm HT}_{\ax}$,
$C^{\rm LT}_{\ax}$ and $\Gm{i}$ are essentially independent on
$m_{\ax}$ (see also \cref{axino}).

In most cases, we observe that the semi-analytical findings
approach rather successfully the numerical ones. Let us clarify,
however, that the results presented in the following sections, are
derived exclusively by our numerical program.

\section{Kination and Gravitino Thermal Abundance}\label{sec:grv}
\setcounter{equation}{0}

We first examine the impact of a KD era on the thermal abundance
of $\Gr$ and study its cosmological consequences. We discriminate
two cases, depending on whether $\Gr$ is unstable
(Sec.~\ref{unGr}) or stable (Sec.~\ref{stGr}).

\subsection{Unstable Gravitino}\label{unGr}

If $\Gr$ is unstable, it can decay after the onset of NS,
affecting the primordial abundances of the light elements in an
unacceptable way. In order to avoid spoiling the success of Big
Bang NS, an upper bound on $Y_{\Gr}$ is to be extracted as a
function of $m_{\Gr}$ and the hadronic branching ratio of $\Gr$,
$B_{\rm h}$ \cite{kohri, kohri2, oliveg}. Let us specify some
representative values of this constraint, taking into account the
most up-to-date analysis of Ref.~\cite{kohri}. In particular, if
$\Gr$ decays mainly to photon and photino, from Fig. 1 of
Ref.~\cite{kohri} we can deduce:
\beq Y_{\Gr}(\Tns)\lesssim\left\{\matrix{
10^{-15} \cr
10^{-14}\cr
10^{-13} \cr}
\right.~~\mbox{for}~~ m_{\Gr}\simeq \left\{\matrix{
360~{\rm GeV} \cr
600~{\rm GeV} \cr
11~{\rm TeV} \cr}\right.~~\mbox{and}~~B_{\rm h}=0.001, \eeq
whereas if $\Gr$ decays mainly to gluons and gluinos, from Fig. 2
of Ref.~\cite{kohri} we can deduce:
\beq Y_{\Gr}(\Tns)\lesssim\left\{\matrix{
10^{-15} \cr
10^{-16} \cr
9.6\times10^{-15} \cr}
\right.~~\mbox{for}~~ m_{\Gr}\simeq \left\{\matrix{
200~{\rm GeV}\hfill \cr
680~{\rm GeV}\hfill \cr
10~{\rm TeV} \hfill \cr}\right.~~\mbox{and}~~B_{\rm h}=1.\eeq
We observe that for $B_{\rm h}=1$, the upper bound on
$Y_{\Gr}(\Tns)$ does not exclusively increase with an increase of
$m_{\Gr}$, as in the case for $B_{\rm h}=0.001$.

In the SC (where no late-time entropy production is expected),
setting $M_{1/2}=500~\GeV$, we obtain a stringent upper bound on
$\Ti$:
\begin{equation} \label{bTr} T_{\rm I}\lesssim\left\{\matrix{
2.3\times10^{6}~{\rm GeV}\hfill\cr
4\times10^{7}~{\rm GeV}\hfill \cr
6\times10^{8}~{\rm GeV}\hfill \cr}
\right. \mbox{for}~~ m_{\Gr}\simeq \left\{\matrix{
360~{\rm GeV} \cr
600~{\rm GeV} \cr
11~{\rm TeV} \cr}\right.~~\mbox{and}~~B_{\rm
h}=0.001,~~\mbox{or}\end{equation}
\begin{equation} \label{bTrh}  T_{\rm I}\lesssim\left\{\matrix{
8.5\times10^{5}~{\rm GeV}\hfill\cr
3.1\times10^{5}~{\rm GeV}\hfill \cr
5.4\times10^{7}~{\rm GeV}\hfill \cr}
\right. \mbox{for}~~ m_{\Gr}\simeq \left\{\matrix{
200~{\rm GeV} \cr
680~{\rm GeV} \cr
10~{\rm TeV} \cr}\right.~~\mbox{and}~~B_{\rm h}=1.\end{equation}
Clearly the upper bound on $T_{\rm I}$ becomes significantly more
restrictive for large $B_{\rm h}$'s and low $m_{\Gr}$'s.

\begin{figure}[t]\vspace*{-.3in}
\begin{center}
\epsfig{file=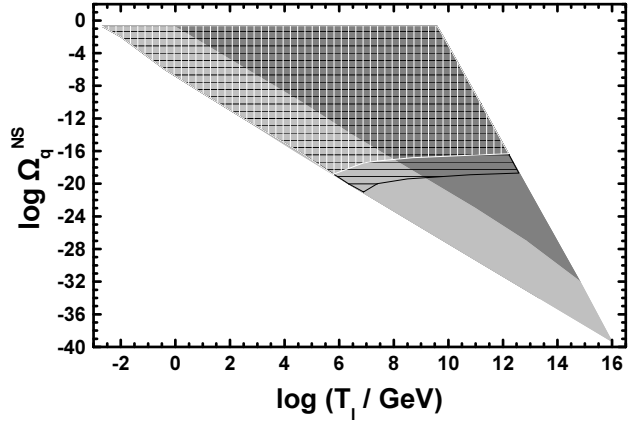,height=3.65in,angle=-90}
\end{center}
\hfill \caption[]{\sl Areas in the $\log\Ti-\log\Omqns$ plane that
are simultaneously allowed by the quintessential requirements
(gray and lightly gray shaded area) -- Eqs.~(\ref{domk}),
(\ref{nuc}) and (\ref{para}) -- and the gravitino constraint for
$m_{\Gr}=M_{1/2}=500~{\rm GeV}$ and $B_{\rm h}=0.001$ (black lined
area) or $B_{\rm h}=1$ (white lined area).} \label{OmTgr}
\end{figure}

In the QKS, if we set $T_{\rm I}=10^9~\GeV$ and $M_{1/2}=500~\GeV$
we can obtain a lower bound on $\Omega_q^{\rm NS}$ which can be
transformed to an upper bound on $\Tkr$ via \Eref{Tkr}. In
particular,
\begin{itemize}
\item For $B_{\rm h}=0.001$,
\beq \hspace*{-1.cm}\Omqns\gtrsim\left\{\matrix{
6\times10^{-19}\hfill\cr
1\times10^{-21}\hfill \cr
6\times10^{-25}\hfill \cr}
\right.\Rightarrow T_{\rm KR}\lesssim\left\{\matrix{
2.8\times10^{5}~{\rm GeV}\hfill \cr
6.8\times10^{6}~{\rm GeV}\hfill \cr
2.8\times10^{8}~{\rm GeV}\hfill \cr}
\right.\mbox{for}~~m_{\Gr}\simeq \left\{\matrix{
360~{\rm GeV}, \hfill \cr
600~{\rm GeV}, \hfill \cr
11~{\rm TeV}. \hfill \cr}
\right.\eeq

\item For $B_{\rm h}=1$,

\beq \hspace*{-1.cm}\Omqns\gtrsim\left\{\matrix{
5\times10^{-18}\hfill\cr
5.5\times10^{-17}\hfill \cr
4.5\times10^{-22}\hfill \cr}
\right.\Rightarrow T_{\rm KR}\lesssim\left\{\matrix{
9.6\times10^{5}~{\rm GeV}\hfill \cr
4.3\times10^{4}~{\rm GeV}\hfill \cr
10^{7}~{\rm GeV}\hfill \cr}
\right.\mbox{for}~ m_{\Gr}\simeq \left\{\matrix{
200~{\rm GeV}, \hfill \cr
680~{\rm GeV}, \hfill \cr
10~{\rm TeV}. \hfill \cr}
\right.\eeq
\end{itemize}

The importance of a KD era in avoiding the gravitino constraint
can also be induced by Fig.~\ref{OmTgr}, where, in contrast to our
previous approach, $\Ti$ is now variable, whereas $m_{\Gr}$ is
fixed to a representative value. In Fig.~\ref{OmTgr}, we show the
regions in the $\log\Ti-\log\Omqns$ plane that are allowed by the
quintessential requirements (see Fig.~\ref{OmT}), for
$m_{\Gr}=M_{1/2}=500~{\rm GeV}$ and $B_{\rm h}=0.001$ (black lined
area) or $B_{\rm h}=1$ (white lined area). We observe that for
$B_{\rm h}=0.001$ the required minimal $\Omqns$ is lower that in
the case of $B_{\rm h}=1$. This is because for $B_{\rm h}=0.001$
we impose $Y_{\Gr}(\Tns)\lesssim2\times10^{-15}$, whereas for
$B_{\rm h}=1$, we impose $Y_{\Gr}(\Tns)\lesssim2\times10^{-16}$
(in accordance with Figs 1 and 2 of Ref.~\cite{kohri}). As a
consequence, the maximal allowed $\Tkr\simeq
(4.8\times10^5-6.8\times10^6)~\GeV$ for $B_{\rm h}=0.001$ is
higher than the one $(4.6\times10^4-5.7\times10^5)~\GeV$ allowed
for $B_{\rm h}=1$. Finally, we observe that the minimal $\Omqns$
depends  very weakly on $\Ti$.

We can conclude, therefore, that the gravitino constraint can be
totally eluded in the QKS, even for tiny values of $\Omega_q^{\rm
NS}$.

\begin{figure}[!t]\vspace*{-.09in}
\hspace*{-.3in}
\begin{minipage}{8in}
\epsfig{file=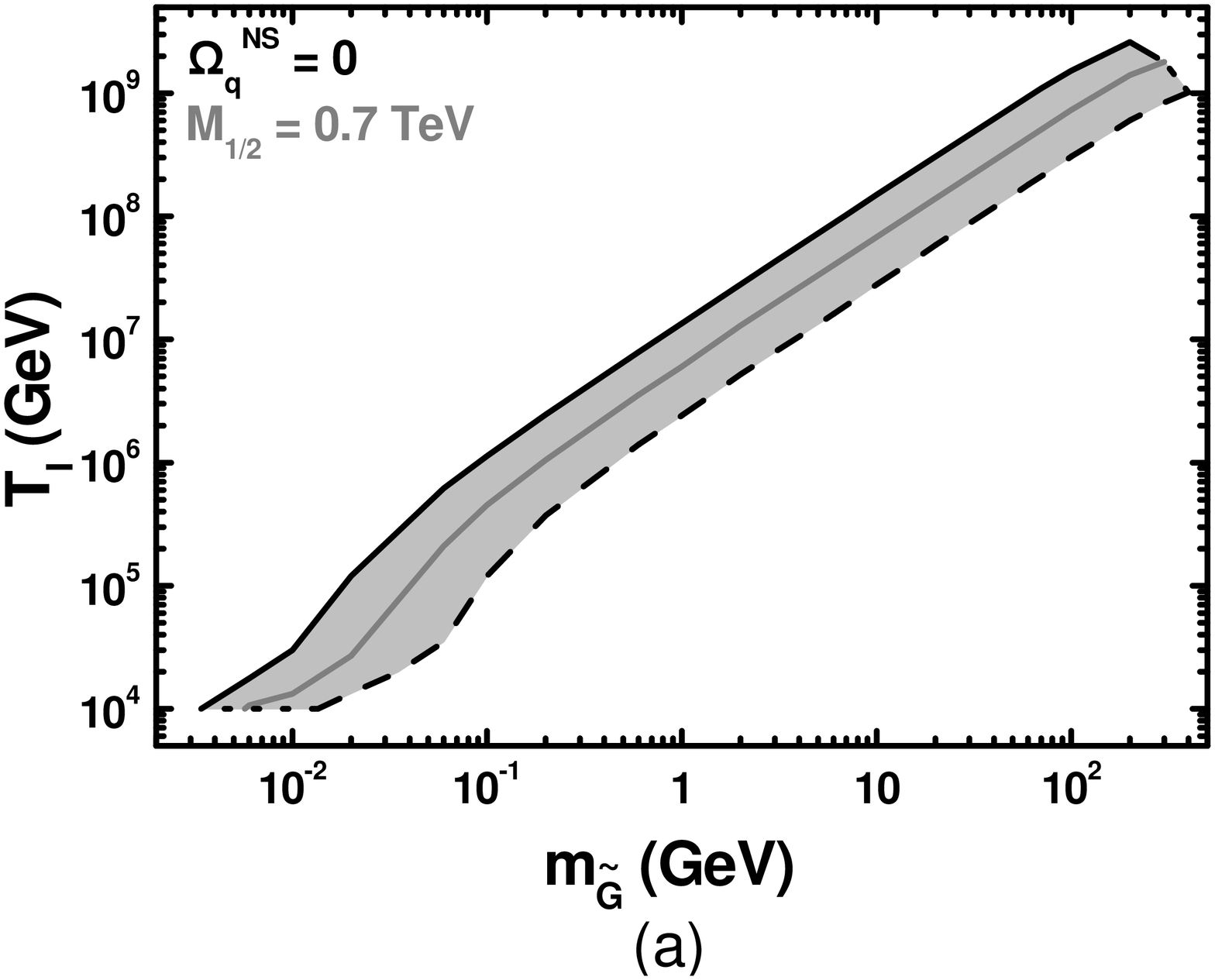,height=3.65in,angle=-90}
\hspace*{-1.37 cm}
\epsfig{file=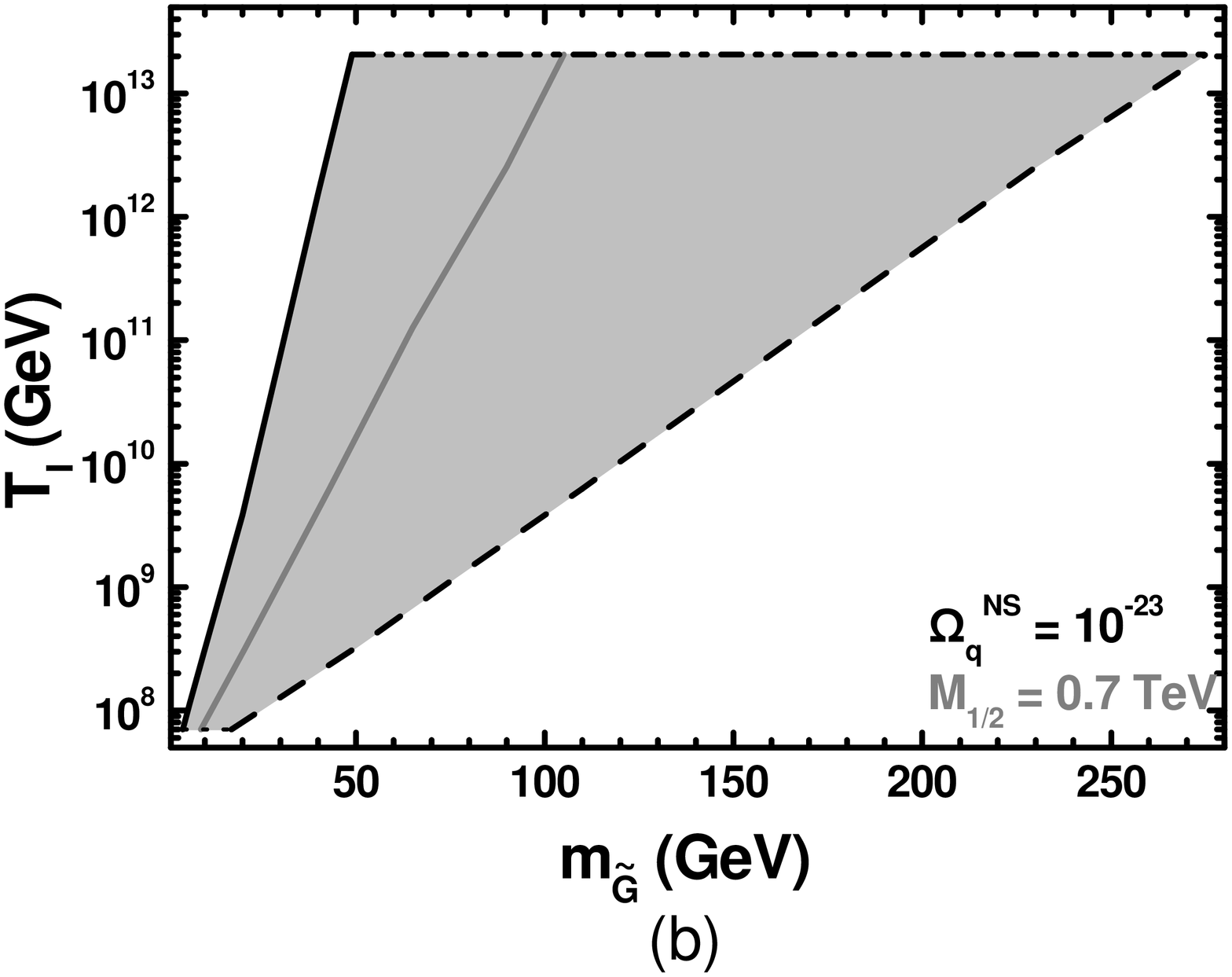,height=3.65in,angle=-90} \hfill
\end{minipage}\vspace*{-.1in}
\hfill\begin{center}
\epsfig{file=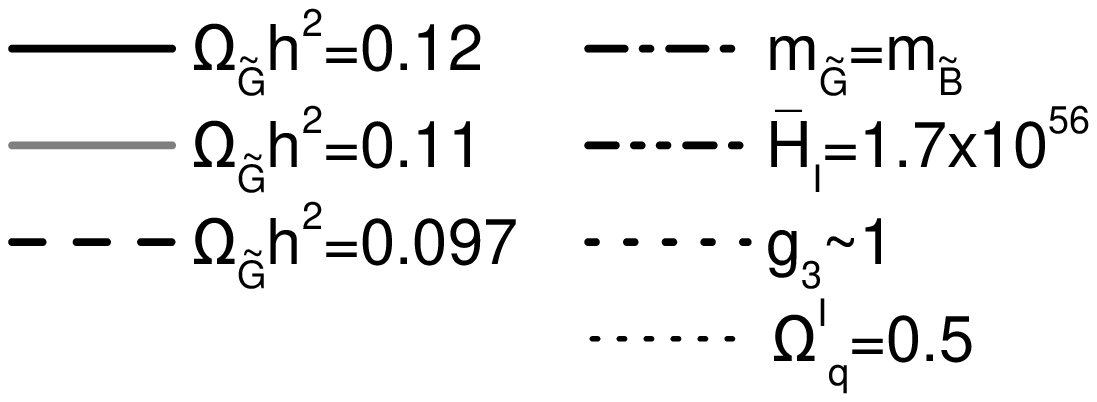,height=2.5in,angle=-90}
\end{center}
\hfill \caption[]{\sl Allowed (lightly gray shaded) regions in the
$m_{\Gr}-\Ti$ plane for $\Gr$-CDM with $0.5\leq M_{1/2}/\TeV\leq1$
in the {\sf\ssz (a)} SC $(\Omqns=0)$ and {\sf\ssz (b)} QKS with
$\Omqns=10^{-23}$. The conventions adopted for the various lines
are also shown.}\label{Tmg}
\end{figure}

\newpage
\subsection{Stable Gravitino}\label{stGr}

$\Gr$ can be stable if it is the LSP within SUSY models with
R-parity conservation. In this case, it constitutes a good CDM
candidate, provided $\Omgr$ satisfies Eq.~(\ref{cdmb}). Imposing
this condition constrains the free parameters, which in this case
are $m_{\Gr}, \Ti, M_{1/2}$ and $\Omqns$.

In \Fref{Tmg} we present the allowed regions (lightly gray shaded)
in the $m_{\Gr}-\Ti$ plane in the {\sf\small (a)} SC $(\Omqns=0)$
and {\sf\small (b)} QKS with $\Omqns=10^{-23}$, letting $M_{1/2}$
vary in the interval $(0.5-1)~\TeV$. The black solid [dashed]
lines correspond to the upper [lower] bound on $\Omgr$ in
\sEref{cdmb}{a}, whereas the gray solid lines have been obtained
by fixing $\Omgr$ to its central value in \sEref{cdmb}{a} for
$M_{1/2}=0.7~\TeV$. In practice, the solid [dashed] line is
constructed for $M_{1/2}=0.5~\TeV$ [$M_{1/2}=1~\TeV$]. This is
because $\Omgr\propto \Ti\,M^2_{\alpha}/m_{\Gr}$ [$\Omgr\propto
\ln\Ti\,M^2_{\alpha}/m_{\Gr}$] in the SC [QKS] as deduced by
Eqs.~(\ref{YhT}) and (\ref{sig1}).

The upper boundary curve (dot-dashed line) in \sFref{Tmg}{a}
results from the requirement that $\Gr$ is the LSP and thus is
lighter than $\tilde B$. Note that $m_{\tilde B}$ varies from
$205$ to $411~\GeV$ for $M_{1/2}$ in the interval $(0.5-1)~\TeV$.
On the other hand, the upper boundary curve (double dot-dashed
line) in \sFref{Tmg}{b} comes from the upper bound on $\vHi$ in
\Eref{para}. The lower boundary curve (dotted [thin] line) in
\sFref{Tmg}{a} [\sFref{Tmg}{b}] arises from the saturation of
$g_3<1$ [\Eref{domk}]. Note that towards lower $T$'s the preferred
$m_{\Gr}$ tends to almost unnaturally small values. We conclude
therefore, that the LTR is not cosmologically interesting in the
case of $\Gr$-CDM.

From \sFref{Tmg}{a} we observe that in the SC ($\Omqns=0$) somehow
larger $\Ti$'s than in the case with unstable $\Gr$ are allowed
(compare, e.g., with \Eref{bTr}). Much larger $\Ti$'s and
$m_{\Gr}$'s are allowed in the QKS -- see \sFref{Tmg}{b}. However,
in this case, we are obliged to fine tune $\Omqns$ to a very low
value $10^{-23}$ $(\Tkr=7\times10^7~\GeV)$, in order to obtain
acceptable $\Omgr$. We consider such a low $\Omqns$ as unnatural
and thus, conclude that $\Gr$ is not a good CDM candidate within
the QKS.

\section{Kination and Axino Thermal Abundance}\label{sec:axn}
\setcounter{equation}{0}

We now turn to $\ax$, as the main candidate for CDM in the
universe. The production of $\ax$ is usually accompanied by the
production of a scalar SUSY partner, the s-axion; this may undergo
out-of-equilibrium decays \cite{saxion}, producing entropy and
therefore, diluting any prexisting $\Omax$. In our analysis, we
assume that this is not the case \cite{axino}. Moreover, we limit
ourselves to $\Ti\leq f_a$ since for larger $\Ti$'s the PQ
symmetry \cite{pq} is restored and so, no particle from the axion
supermultiplet has been produced.

Let us initially derive the temperature $T_{\rm D}$ at which $\ax$
decouples from the thermal bath in the SC and QKS. The
$\ax$-decoupling occurs when
\beq \label{TD} H(T_{\rm D})\simeq\Gamma_{\ax}(T_{\rm
D}),~~\mbox{where}~~\Gamma_{\ax}=\sigv{\ax}\, n^{\rm eq}\sim
6\,N_{\rm F}(N_3^2-1)g_a^2g_3^2\, n^{\rm eq}/2\eeq
is an update \cite{axinold} of the interaction rate which
maintains $\ax$'s in chemical equilibrium with the thermal bath
(we take $N_{\rm F}=12$ and $N_3=3$ as explained in Appendix
A). Similarly to the previous discussion, we distinguish two cases:

\begin{itemize}
\item In the SC, solving \Eref{TD} w.r.t $T_{\rm D}$ (after
replacing $H$ by its expression in \Eref{eq3} for $r_q=0$), we
find
\bea \label{TDsc} T^{\rm SC}_{\rm D}\simeq{\pi^4g^{\rm D}_*\over
8640\sqrt{3}m_{\rm
P}g_a^2g_3^2\zeta(3)}\simeq10^8~\GeV~~\mbox{for}~~f_a=10^{11}~\GeV~~\mbox{and}\\
6\times10^5\lesssim T^{\rm SC}_{\rm D}/\GeV\lesssim
2\times10^{10}~~\mbox{for}~~10^{10}\leq
f_a/\GeV\leq10^{12}.\label{TDsc1}\eea

\item In the QKS (after replacing $H$ in \Eref{TD} by its expression in
\Eref{eq3} with $r_q$ given by \Eref{rqap}), we find
\bea \label{TDq} H>\Gamma_{\ax}~~\mbox{for}~~\Tkr<T^{\rm SC}_{\rm
D}~~\mbox{or}~~\Omqns\gtrsim{g_*^{\rm NS}\over g_*^{\rm
D}}\left({\Tns\over T^{\rm SC}_{\rm
D}}\right)^2\\\label{TDq1}\Rightarrow~~\Omqns\gtrsim\left\{\matrix{
1\times10^{-19}&\mbox{for}~~f_a=10^{10}~\GeV,\cr
3.5\times10^{-24}&\mbox{for}~~f_a=10^{11}~\GeV,\cr
1\times10^{-28} &\mbox{for}~~f_a=10^{12}~\GeV.  \cr}
\right.\eea
These values were extracted numerically without the
approximation of \Eref{rqap} and therefore, are less restrictive
that the result of the analytical estimate in \Eref{TDq}. Note
that the constraints on $\Omqns$ are independent on $\Ti$.

\end{itemize}

As we already emphasize in \Sref{sec:boltz} we expect that the
primordial (i.e., due to the $\ax$-decoupling) $\ax$ yield
\cite{steffenaxino} ($Y^{\rm D}_{\ax}=(3n^{\rm eq}/2s)(T_{\rm
D})\simeq1.8\times10^{-3}$) is diluted by the entropy release
during reheating of the universe to a temperature $\vTi$. Under
this assumption, we have to ensure that $\Ti<T^{\rm SC}_{\rm D}$
[$\Tkr<T^{\rm SC}_{\rm D}$] constructing the regions where $\ax$
is a viable CDM-candidate in the SC [QKS].

Our results for the HTR and the LTR are analyzed separately in the
following. Let us remind that the discrimination between the two
regimes (HTR or LTR) depends on the hierarchy not only between
$\Ti$ and $\Tc$ but also between $\Tkr$ and $\Tc$. In the SC we
have $\Ti\gg\Tc$ [$\Ti\ll\Tc$] for the HTR [LTR] (see \Sref{hT}
[\Sref{hT}]). In the QKS we take $\Ti\gg\Tc$ and $\Tkr>\Tc$ for
the HTR (see \Sref{hT}), but we consider as more natural choice
(motivated by the majority \cite{susyhybrid} of the inflationary
models) to take $\Ti\gg\Tc$ and $\Tkr<\Tc$ for LTR (see
\Sref{lT}). The dependence of our results on the variation of
$\Ti$ in the QKS is also studied in \Sref{lT}.

\subsection{The High $T$ Regime}\label{hT}

\begin{figure}[!t]\vspace*{-.09in}
\hspace*{-.3in}
\begin{minipage}{8in}
\epsfig{file=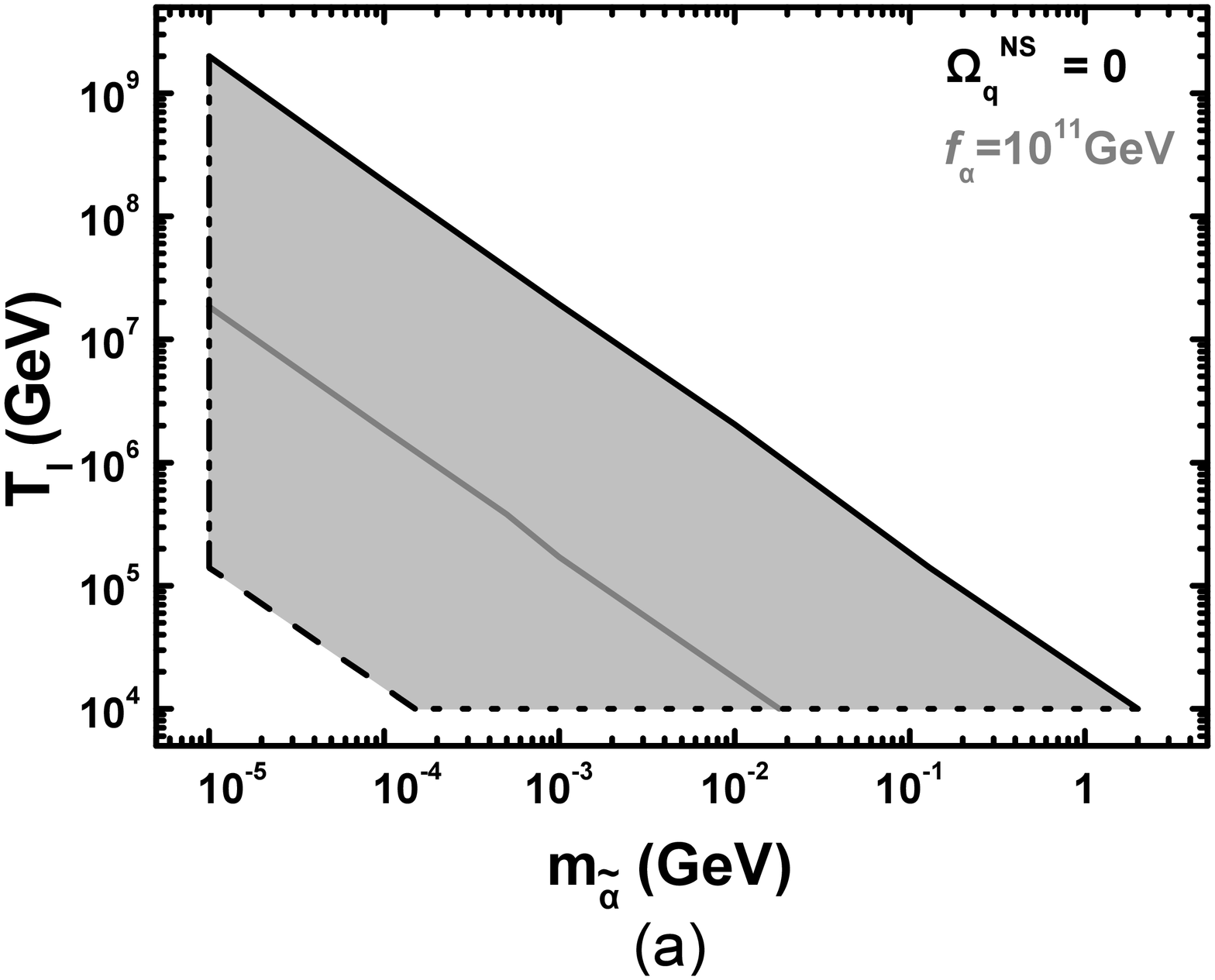,height=3.65in,angle=-90}
\hspace*{-1.37 cm}
\epsfig{file=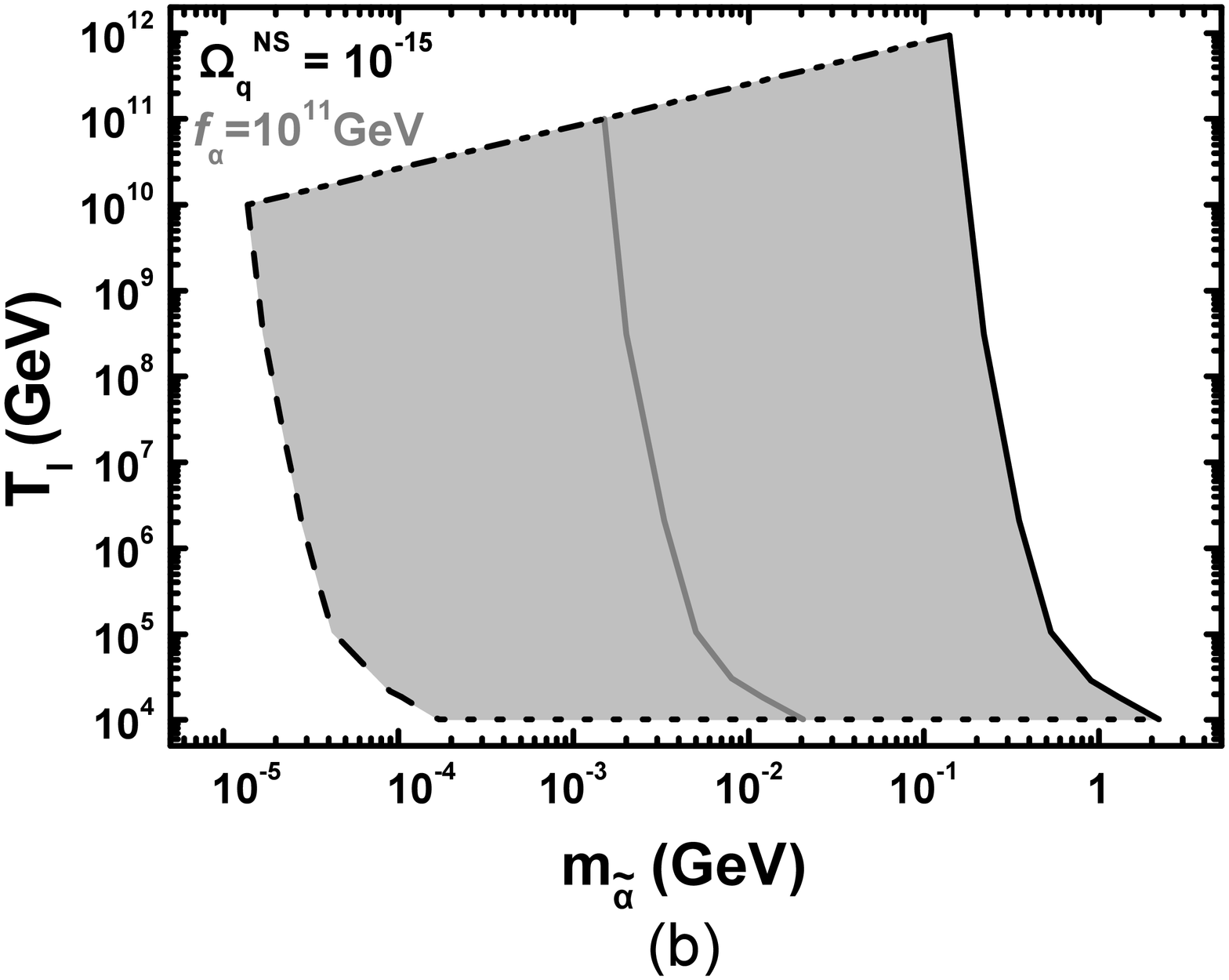,height=3.65in,angle=-90} \hfill
\end{minipage}\vspace*{-.1in}
\hfill\begin{center}
\epsfig{file=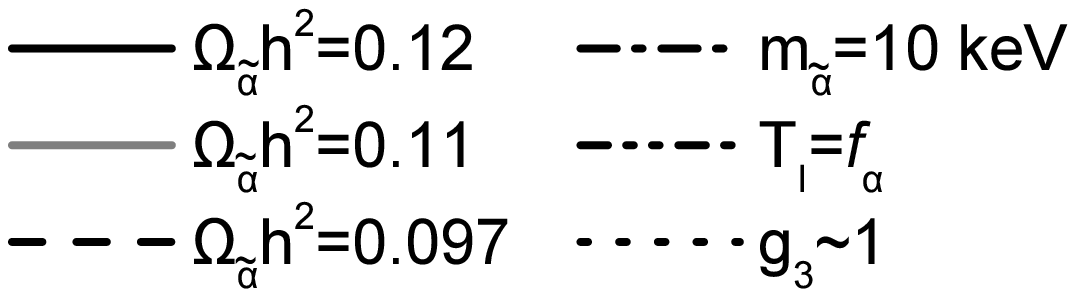,height=2.5in,angle=-90}
\end{center}
\hfill \caption[]{\sl Allowed (lightly gray shaded) regions for
$\ax$-CDM in the HTR and $m_{\ax}-\Ti$ plane with $10^{10}\leq
f_a/\GeV\leq10^{12}$ for the {\sf\ssz (a)} SC ($\Omqns=0$) and
{\sf\ssz (b)} QKS with $\Omqns=10^{-15}$. The conventions adopted
for the various lines are also shown.}\label{Tma}
\end{figure}

In this case the HTLA works well and therefore the comparison with
the case of $\Gr$-CDM is straightforward; the free parameters
in the present case are: $m_{\ax},~f_{a},~\Ti$ and
$\Omqns$.

In \Fref{Tma} we display the allowed regions (lightly gray shaded)
in the $m_{\ax}-\Ti$ plane in the {\sf\small (a)} SC $(\Omqns=0)$
and {\sf\small (b)} QKS with $\Omqns=10^{-15}$, letting $f_a$ vary
in the interval $(10^{10}-10^{12})~\GeV$. The selected $\Omqns$ is
the largest possible value that ensures the validity of HTLA,
since $\Tkr\simeq10^4~\GeV$. The black solid [dashed] lines
correspond to the upper [lower] bound on $\Omax$ in
\sEref{cdmb}{a}, whereas the gray solid lines have been obtained
by fixing $\Omax$ to its central value in \sEref{cdmb}{a} for
$f_a=10^{11}~\GeV$. In practice, the solid [dashed] line is
constructed for $f_a=10^{12}~\GeV$ [$f_a=10^{10}~\GeV$]. This is
because $\Omax\propto \Ti\,m_{\ax}/f^2_{a}$ [$\Omax\propto
\ln\Ti\,m_{\ax}/f^2_{a}$] in the SC [QKS] as deduced by
Eqs.~(\ref{YhT}) and (\ref{sig1}).

The left boundary curve (dot-dashed line) in \Fref{Tma}-{\sf\small
(a)} comes from the lower bound of \sEref{cdmb}{b}. On the other
hand, the upper boundary curve (double dot-dashed line) in
\sFref{Tma}{b} comes from the upper bound on $\Ti$, $\Ti\leq f_a$.
The lower boundary curves (dotted lines) in both \sFref{Tma}{a}
and \sFref{Tma}{b} come from the saturation of $g_3>1$ which
determines the range of validity of the HTLA. Needless to say
that, for $\Ti<10^4~\GeV$, imposing \Eref{domk} also fails, since
$\Ti\simeq\Tkr$.

We observe that in the SC the lower bound of \sEref{cdmb}{b} is
more restrictive than our requirement $\Ti<T^{\rm SC}_{\rm D}$,
which is satisfied for all $f_a$'s -- see Eqs.~(\ref{TDsc}) and
(\ref{TDsc1}). On the other hand, in the QKS, the selected
$\Omqns$ fulfills \Eref{TDq1}. For central values of $f_a$ and
$\Omax$ we find $1.85\times10^7\lesssim\Ti/\GeV\lesssim10^4$ for
$10^{-5}\lesssim m_{\ax}/\GeV\lesssim0.018$ in the SC, whilst
$10^{11}\lesssim\Ti/\GeV\lesssim10^4$ for $0.0015\lesssim
m_{\ax}/\GeV\lesssim0.02$ in the QKS. We observe that in the SC
the allowed $\Ti$ and $m_{\ax}$ are rather low, whereas in the QKS
larger $\Ti$ and $m_{\ax}$ are permitted. Note, also, that towards
lower $\Ti$ the preferred $m_{\ax}$ tends to larger values,
implying that the LTR is cosmologically interesting in the case of
$\ax$-CDM (contrary to the case of $\Gr$-CDM).

\subsection{The Low $T$ Regime}\label{lT}

\begin{figure}[!t]\vspace*{-.09in}
\hspace*{-.3in}
\begin{minipage}{8in}
\epsfig{file=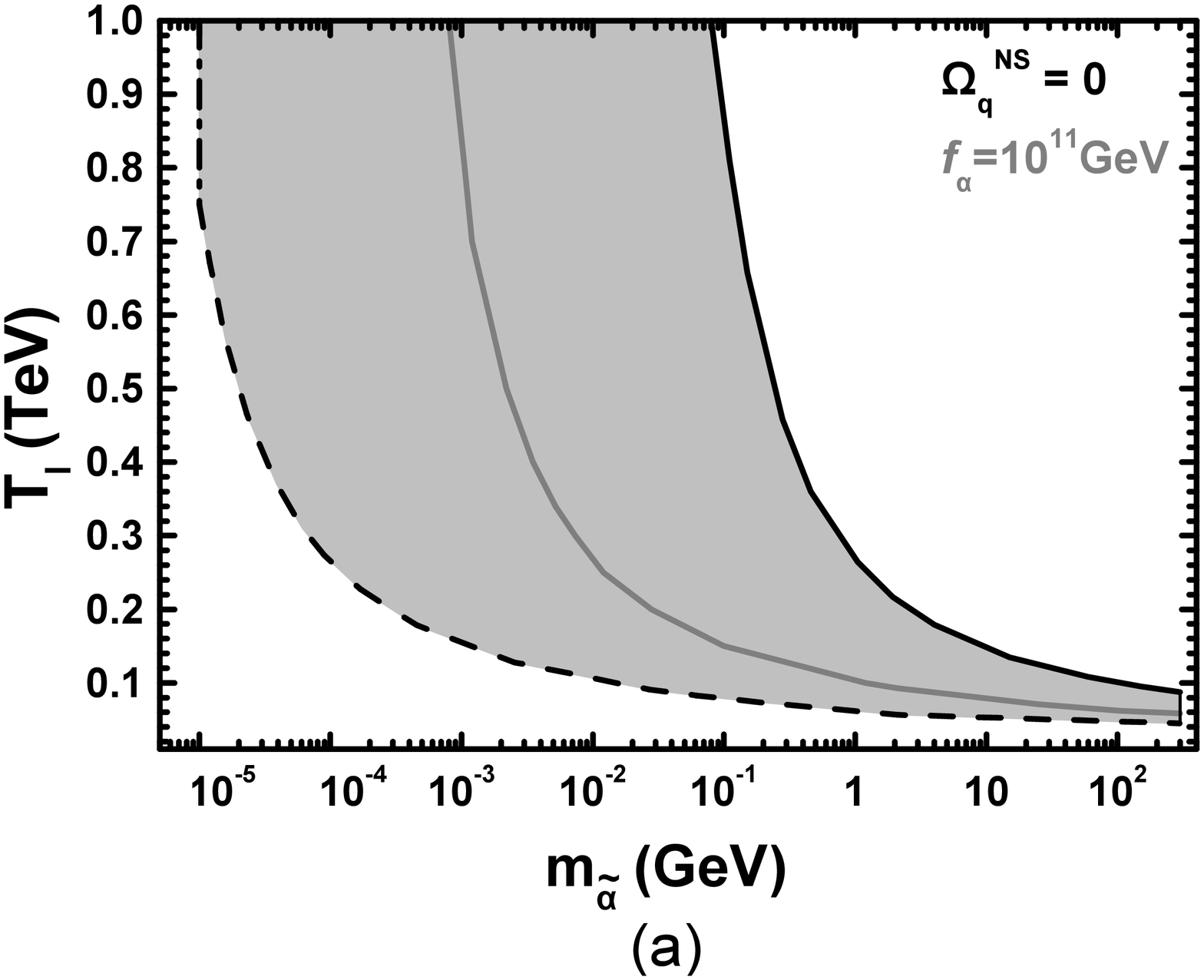,height=3.65in,angle=-90}
\hspace*{-1.37 cm}
\epsfig{file=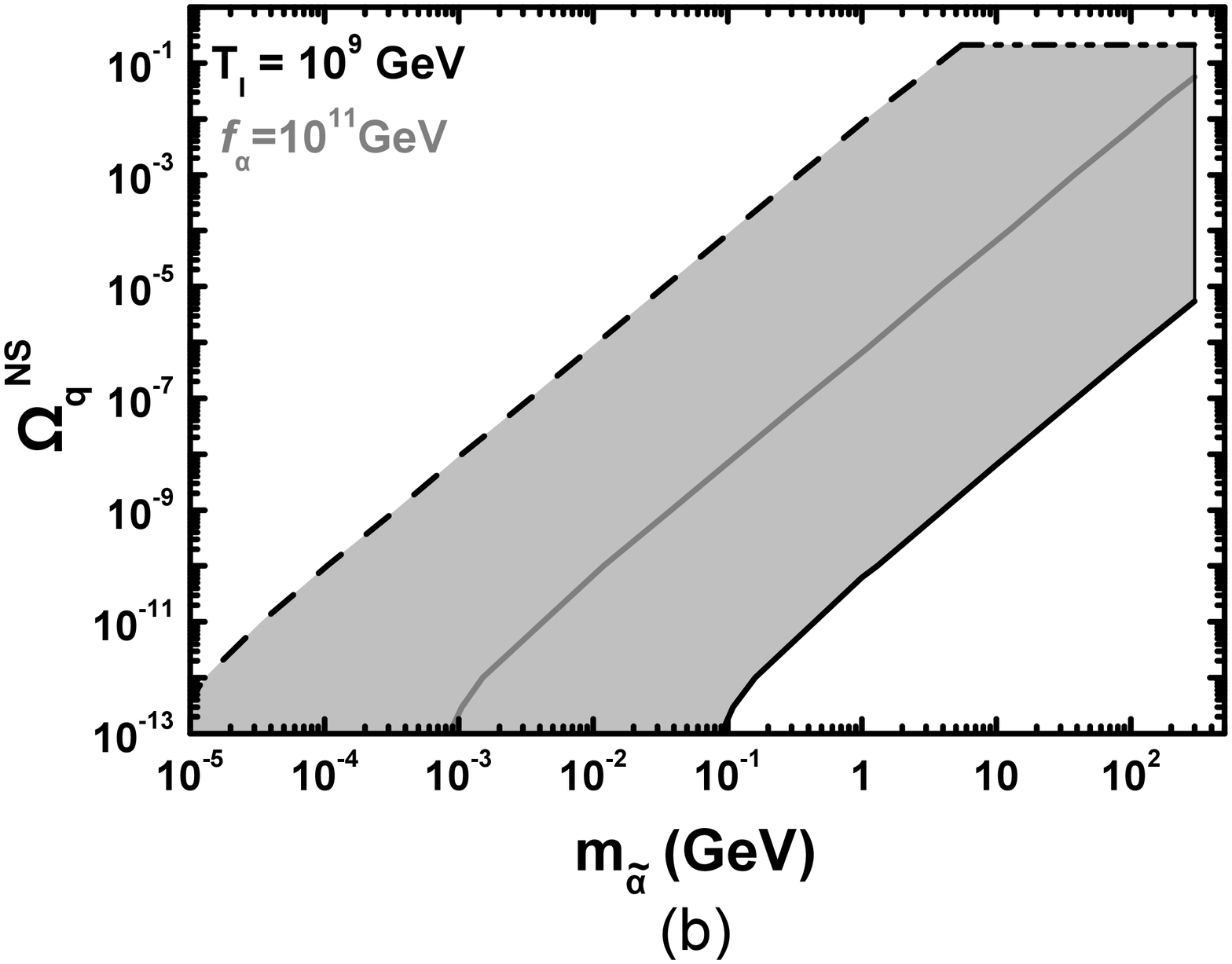,height=3.65in,angle=-90} \hfill
\end{minipage}\vspace*{-.1in}
\begin{center}
\epsfig{file=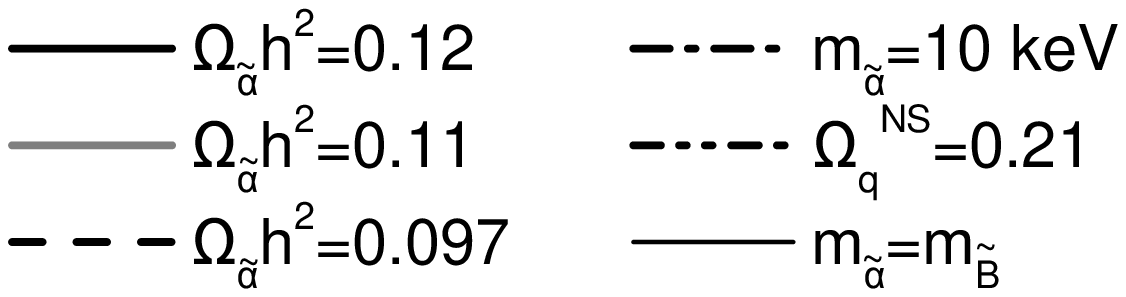,height=2.5in,angle=-90}
\end{center}
\hfill \caption[]{\sl Allowed (lightly gray shaded) regions for
$\ax$-CDM, $10^{10}\leq f_a/\GeV\leq10^{12}$ and $m_i$'s shown in
\Eref{mi} in the LTR and {\sf\ssz (a)} in the $m_{\ax}-\Ti$ plane
for the SC ($\Omqns=0$) and {\sf\ssz (b)} in the $m_{\ax}-\Omqns$
plane for the QKS with $\Ti=10^9~\GeV$. The conventions adopted
for the various lines are also shown.}\label{Tmd}
\end{figure}

In this regime $\Omax$ includes sizeable contributions from
thermal scatterings and decay of SUSY partners. As a consequence,
our computation depends not only on the parameters of the previous
case ($m_{\ax},~f_{a},~\Ti$ and $\Omqns$), but also on the masses
of the SUSY particles, $m_i$. We use the $m_i$'s of \Eref{mi} for
both the SC and the QKS. Needless to say that our preliminary
assumptions $\Ti<f_a$ and $\Ti<T^{\rm SC}_{\rm D}$ [$\Tkr<T^{\rm
SC}_{\rm D}$] are comfortably satisfied in the SC [QKS].

In the SC ($\Omqns=0$), the allowed (lightly gray shaded) region
for $\ax$-CDM is shown  in the $m_{\ax}-\Ti$ plane -- see
\sFref{Tmd}{a}. On the other hand, in the QKS, our results are
independent of $\Ti$ -- see \Eref{emp}; consequently, we depict
the allowed (lightly gray shaded) region in the $m_{\ax}-\Omqns$
plane with fixed $\Ti=10^9~\GeV$ -- see \sFref{Tmd}{b}. The black
solid [dashed] line corresponds to the upper [lower] bound on
$\Omax$ in \sEref{cdmb}{a}, whereas the gray solid lines have been
obtained by fixing $\Omax$ to its central value in \sEref{cdmb}{a}
for $f_a=10^{11}~\GeV$. In practice, the solid [dashed] line is
constructed for $f_a=10^{12}~\GeV$ [$f_a=10^{10}~\GeV$]. This is
because $\Ti\propto B\ln(f_a^2\Omax)-B\ln(A\,m_{\ax})$
[$\sqrt{\Omqns}\propto m_{\ax}/\Omax f^2_{a}$] in the SC [QKS] as
deduced by \Eref{emp0} [\Eref{emp}].

In the SC, the lower bound of \sEref{cdmb}{b} determines a part
(dot-dashed line) of the left boundary curve in \sFref{Tma}{a},
whereas the upper bound of \sEref{cdmb}{b} sets an upper bound
(thin lines) in both \sFref{Tma}{a} and \sFref{Tma}{b}. The upper
boundary curve (double dot-dashed line) in \sFref{Tma}{b} comes
from the upper bound on $\Omqns$ in \Eref{nuc}. The allowed area
in \sFref{Tma}{a} [\sFref{Tma}{b}] terminates from above [below]
at $\Ti\simeq1~\TeV$ [$\Tkr\simeq 1~\TeV$], so that our formulas
for $C^{\rm LT}_{\ax}$ to be fully applicable.

A sharp suppression of $\Omax$ is observed, due to the Boltzmann
suppression factor $e^{-B/\Ti}$, in the SC -- see \sFref{Tmd}{a}.
In this case, larger $m_{\ax}$ but very low $\Ti$ are allowed. On
the contrary, $\Ti$ can be fixed to a naturally \cite{susyhybrid}
high value within the QKS. At the same time, $m_{\ax}$ and
$\Omqns$ take interestingly large values in the allowed region --
see \sFref{Tmd}{b}. For these reasons, this case is considered as
the most intriguing of this paper. For central values of $f_a$ and
$\Omax$ we find $58.5\lesssim\Ti/\GeV\lesssim1000$ for $300\gtrsim
m_{\ax}/\GeV\gtrsim8\times10^{-4}$ in the SC, whereas
$10^{-13}\lesssim\Omqns\lesssim0.056$ for
$8.5\times10^{-4}\lesssim m_{\ax}/\GeV\lesssim300$ in the QKS
($\Ti=10^9~\GeV$). Our formalism of Appendix A is of crucial
importance in order to obtain a reliable result in both the SC and
QKS.

\begin{figure}[!t]\vspace*{-.3in}
\begin{center}
\epsfig{file=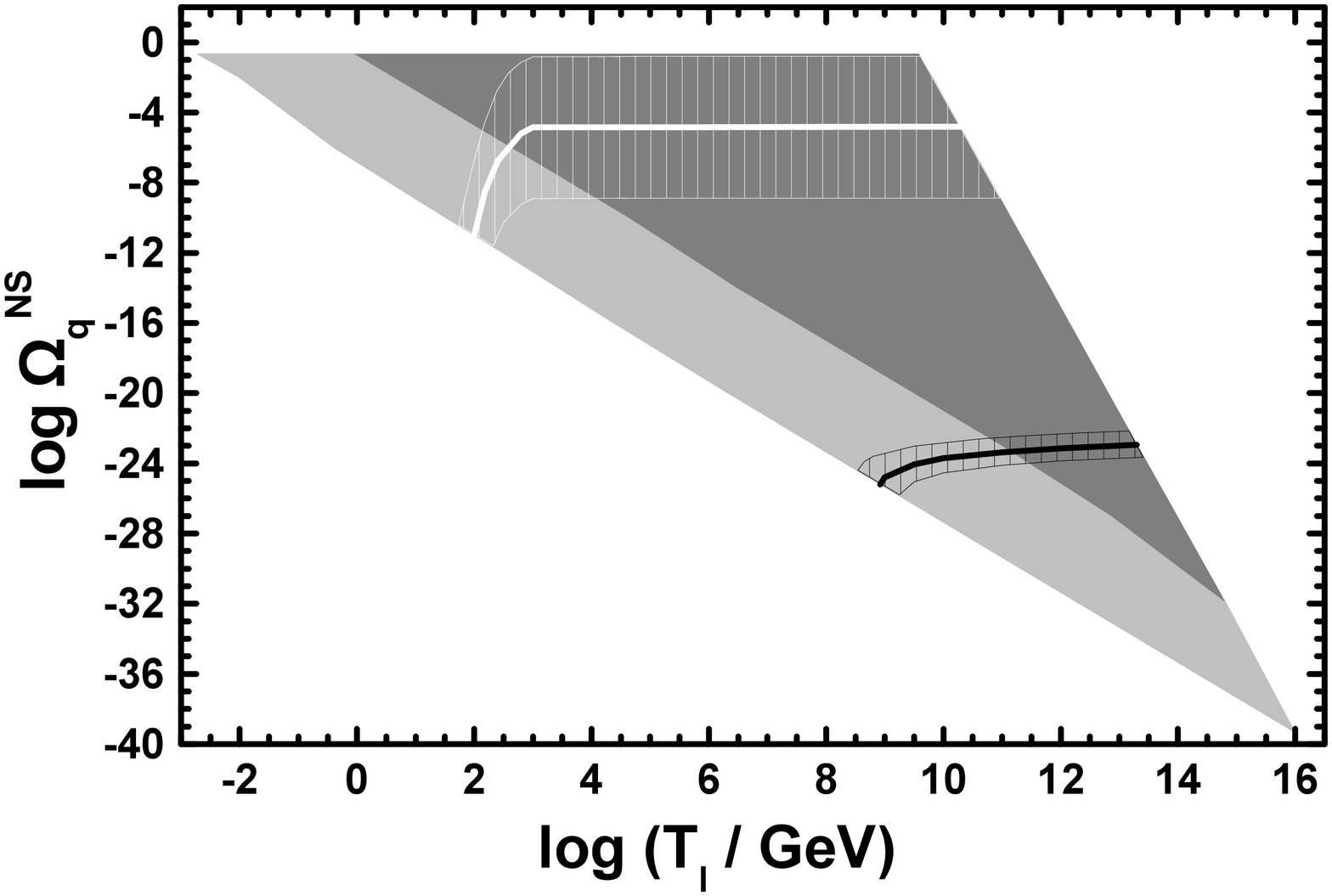,height=3.65in,angle=-90}
\end{center}
\hfill \caption[]{\sl Simultaneously allowed areas in the
$\log\Ti-\log\Omqns$ plane taking into account the quintessential
requirements (gray and lightly gray shaded area) --
Eqs.~(\ref{domk}), (\ref{nuc}) and (\ref{para}) -- and the CDM
constraint for $\Gr$-CDM (black lined area) with
$m_{\Gr}=100~\GeV$ and $0.5\leq M_{1/2}/\TeV\leq1$ or $\ax$-CDM
(white lined area) with $m_{\ax}=5~\GeV$, $m_i$'s of \Eref{mi} and
$10^{10}\leq f_a/\GeV\leq10^{12}$ (white lined area). The bold
black [white] line is obtained for $M_{1/2}=0.7~\TeV$
[$f_a=10^{11}~\GeV$] and corresponds to $\Omx\simeq0.11$ (with
$X=\Gr$ [$X=\ax$]).} \label{OmTcdm}
\end{figure}

Finally, it would be interesting to directly compare the
naturality of the $\ax$ and $\Gr$ as CDM candidates in the QKS.
This is done in \Fref{OmTcdm} where, contrary to our strategy in
\sFref{Tmg}{b} [\sFref{Tmd}{b}], we fix $m_{\Gr}$ [$m_{\ax}$] to
some exemplary value and let $\Ti$ and $\Omqns$ vary in their
allowed region of \Fref{OmT}. In particular, in \Fref{OmTcdm} we
present in the $\log\Ti-\log\Omqns$ plane the allowed region by
both the quintessential requirements and the CDM constraint for
$\Gr$-CDM or $\ax$-CDM. The first set of constraints --
Eqs.~(\ref{domk})-(\ref{para}) -- is satisfied in the gray and
lightly gray shaded area. The CDM constraint -- \Eref{cdmb} -- for
$\Gr$-CDM with $m_{\Gr}=100~\GeV$ and $0.5\leq M_{1/2}/\TeV\leq1$
is fulfilled in the black lined region, where $\Omqns$ is tuned to
low values. The upper [lower] boundary curve of the black lined
region corresponds to $\Omgr=0.097$ [$\Omgr=0.12$] and is
constructed for $M_{1/2}=1~\TeV$ [$M_{1/2}=0.5~\TeV$]. This is
because $\sqrt{\Omqns} \propto M^2_\alpha/m_{\Gr}\Omgr$ as
concluded by Eqs.~(\ref{sig1}) and (\ref{YhT}). On the contrary,
\Eref{cdmb} for $\ax$-CDM with $m_{\ax}=5~\GeV$ and $10^{10}\leq
f_{a}/\GeV\leq10^{12}$ is met in the white lined region, with much
more natural $\Omqns$'s than the ones needed in the black lined
region. For the required $\Omqns$'s our results of the LTR are
applicable. For $\Ti\gtrsim 1~\TeV$, $\Omax$ is obviously
independent on $\Ti$ and \Eref{emp} approaches fairly our
numerical results. The upper [lower] boundary curve of the white
lined region corresponds to $\Omax=0.097$ [$\Omax=0.12$] and is
constructed for $f_a=10^{10}~\GeV$ [$f_a=10^{12}~\GeV$] (in
accordance with our discussion above). The bold black [white]
lines are constructed for $M_{1/2}=0.7~\TeV$ [$f_a=10^{11}~\GeV$]
and correspond to $\Omx\simeq0.11$. For these values, we find
$8\times10^8\lesssim\Ti/\GeV\lesssim2\times10^{13}$ and
$6.3\times10^{-26}\lesssim\Omqns\lesssim10^{-23}$ for $\Gr$-CDM,
whereas $92.2\lesssim\Ti/\GeV\lesssim1.8\times10^{10}$ and
$1.25\times10^{-11}\lesssim \Omqns\lesssim1.6\times10^{-5}$ for
$\ax$-CDM. Therefore, in the framework of the QKS, $\ax$ is
clearly a more natural CDM candidate than $\Gr$.

\section{Conclusions}\label{sec:con}

We presented an exponential quintessential model which generates a
period dominated by the kinetic energy of the quintessence field.
The parameters of the quintessential model ($\lambda, \Ti,
\Omqns$) were confined so that $0.5\leq\Omega_q(\Ti)\leq1$, and were
constrained by current observational data originating from
NS, the acceleration of the universe, the inflationary scale and
the DE density parameter. We found $0<\lambda<0.9$ and studied the
allowed region in the ($\Ti, \Omqns$)-plane.

We proceeded to examine the impact of this KD epoch to the thermal
abundance of $\Gr$ and $\tilde a$. We solved the problem
{\sf\small (i)} numerically, integrating the relevant system of
the differential equations equation that governs the evolution of
the $\chia$-number density and {\sf\small (ii)} semi-analytically,
producing approximate relations for the cosmological evolution
before and after the transition from KD to RD, and solving the
appropriately re-formulated Boltzmann.  Although we did not
succeed to achieve general analytical solutions in all cases, we
consider as a significant development the derivation of a result
by solving numerically just one equation, instead of the whole
system above. Moreover, for typical values of $m_i$'s in
\Eref{mi}, empirical formulas that reproduce quite successfully
our numerical results were derived. For unstable $\Gr$, the
$\Gr$-constraint poses a lower bound on $\Omqns$, which turns out
to be almost independent of $\Ti$. The CDM constraint can be
satisfied by the thermal abundance of $\Gr$ for extremely low
values $(10^{-23}-10^{-24})$ of $\Omqns$. On the contrary, the
former constraint can be fulfilled by the thermal abundance of
$\ax$ with values of $\Omqns$ close to the upper bound posed by
the requirement for successful NS.

Let us also comment here on three minor subtleties of our
calculation which, do not alter the basic features of our
conclusions (although could potentially create some quantitative
modifications to our results). In particular:

\begin{itemize}

\item Throughout our investigation we did not identify the nature
of NLSP. Therefore the upper bound, shown in \sFref{Tmg}{a} [Fig.
7], on $m_{\Gr}$ [$m_{\ax}$] derived from the requirement
[$m_{\Gr}\leq m_{\tilde B }$] $m_{\ax}\leq m_{\tilde B }$ could be
modified if there is another SUSY particle lighter than $\tilde
B$. Moreover, we did not consider the NS constraints concerning
the late decays of the NLSP into $X$'s. These constraints
\cite{axino,gravitinont, axinont} depend very much on the
properties of the NLSP, i.e. its composition, its mass relative to
the $m_X$, and its coupling to $X$'s. Consequently, additional
bounds on the $m_X$ might arise. As the $\ax$ interactions are not
as strongly suppressed as the $\Gr$ interactions, the $\ax$ LSP
anyhow is far less problematic than the $\Gr$ LSP w.r.t these
constraints.

\item In the case of $\Gr$, we did not incorporate contributions
to $\Omgr$ from the process of reheating. Indeed, these extra
contributions can be  a fraction of the result shown in \Eref{YhT}
\cite{kohri2,sahu}, in the case of the usual reheating realized by
the coherent oscillations of a massive particle \cite{turner}.
However, in the QKS several reheating processes have been proposed
\cite{sami} and therefore, any  safe comparison between the SC and
the QKS has to be performed for $T<\Ti$. In other words, to keep
our investigation as general as possible, we preferred to study
the evolution of the universe after the start of the RD [KD] era
in the SC [QKS] (we simply assumed the existence of an earlier
inflationary epoch).

\item  In the case of the $\ax$-CDM, we used throughout our
investigation some representative masses for the superpartners,
given in \Eref{mi}. Variation in these values (especially in
$m_{\gl}$ and $m_{\sq}$) has an impact on $\Omax$ in the LTR
(e.g., for $m_{\gl}\simeq m_{\sq}$ the contribution of the process
$\sq^*q$ to $C_{\ax}^{\rm LT}$ can be enhanced). In addition, a
further uncertainty in our calculation arises from the
determination of $\Ts$ below which $C_{\ax}^{\rm HT}$ is replaced
by $C_{\ax}^{\rm LT}$ in the integration of the relevant
equations. Note however, that the aim of the present paper is to
demonstrate the change of $\Omx$ due to the presence of the KD era
and not a full scan of the SUSY parameter space.
\end{itemize}

Although our results have been derived in the context of an
exponential quintessential model, their applicability can be
extended to every model that generates a KD phase, even without
\cite{kination} quintessential consequences. It is worth
mentioning that in the presence of kination we can obtain
simultaneous compatibility of both the gravitino and the CDM
constraint (i.e., the lower bound on $\Omqns$ from the gravitino
constraint is compatible with the $\Omqns$'s needed in order to
have the correct amount of $\ax$ CDM). It would be probably
interesting to check if we can obtain a simultaneous compatibility
of these two constraints with additional bounds, arising e.g. from
leptogenesis and neutrino masses \cite{kinlept, Bento} or the
quintessino abundance \cite{quintessino}.

\ack We would like to thank K.Y. Choi, R. Ruiz de Austri and L.
Roszkowski for helpful discussions. The research of S.L was funded
by the FP6  Marie Curie Excellence Grant MEXT-CT-2004-014297. The
work of M.E.G, C.P and J.R.Q was supported by the Spanish MEC
project FPA2006-13825 and the project P07FQM02962 funded by the
``Junta de Andalucia''.

\appendix

\section{Axino Production via Scatterings at Low Temperature}
\setcounter{equation}{0}
\renewcommand{\theequation}{A\arabic{equation}}

In this appendix we present the necessary ingredients used for the
evaluation of the $\ax$-production via scatterings at low $T$. Our
starting point is the general formula in \cref{falk}, which gives
the thermal averaged cross section times the relativistic
invariant relative velocity, $\sigv{ij}$, of two particles $i$ and
$j$ with masses $m_i$ and $m_j$ (in general $m_i\neq m_j$) and
degrees of freedom $g_i$ and $g_j$ respectively:
\begin{equation}
\sigv{ij}= {1\over2m_i^2m_j^2 TK_2\left({m_i\over
T}\right)K_2\left({m_j\over T}\right)} \int_{s_0}^\infty
ds\,w_{ij}(s)K_1\left({\sqrt{s}\over T}\right)p_{\rm i}(m_i,m_j)
\; \label{sv0}
\end{equation}
where $s_0=(m_i+m_j)^2$ and
\beq w_{ij}(s) = {g_ig_j \over32\pi}\, {p_{\rm f}(m_k,m_{\ax})
\over\sqrt{s}}\int_{-1}^{+1}d\cos\tcm\,\tsq{ij}\;. \label{wij}
\eeq
Here, $\tsq{ij}$ is the squared amplitude of the process
$i+j\rightarrow k+\ax$ which contributes to $\sigv{ij}$, $m_k$ is
the mass of the outgoing particle $k$ in this process,
$\theta_{\rm CM}$ is the scattering angle in the center-of-mass
frame and $p_{\rm i}$ and $p_{\rm f}$ are the magnitudes of the
incoming and outcoming 3-momentums in the same reference frame;
these are given respectively by:
\bea && p_{\rm i}(m_i,m_j)={\lambdaup(m_i,m_j)\over \sqrt{s}}
~~\mbox{and}~~ p_{\rm f}(m_k,m_{\ax})={\lambdaup(m_k,m_{\ax})\over
\sqrt{s}}\label{ps1}\\
&& \mbox{where}~~\lambdaup^2(m_i,m_k)={1\over
4}\left(s-(m_i+m_k)^2\right)\left(s-(m_i-m_k)^2\right)\label{ps}
\eea
Multiplying $\sigv{ij}$ by $\nequ_i \nequ_j$ and using \Eref{neq}
we arrive at our final result in \Eref{sig}, with $C_\chia=C^{\rm
LT}_{\ax}$ given by \Eref{sig2}.

For the manipulation of this result it would be useful to remember
that $\tsq{ij}$ is in general a function of the Mandelstam
variables, $s,~t$ and $u$ (and the masses of the involved particles).
However, one of the Mandelstam variables, typically $u$,
can be eliminated in favor of the other two, through the formula:
\beq u=m_i^2+m_j^2+m_k^2+m_{\ax}^2-s-t \label{mad}\eeq
whereas in the reference frame of the center of mass, $t$ can be
expressed in terms of $s$ and $\theta_{\rm CM}$ using \Eref{mad}
and the relation \cite{falk}:
\beq t-u=-{(m^2_{i}-m^2_{j}) (m_k^2-m_{\ax}^2)\over s}+4 p_{\rm
i}(s) p_{\rm f}(s)\cos\theta_{\rm CM}\label{tu}\eeq
Therefore, $\tsq{ij}$ can be written as a function of only $s$ and
$\theta_{\rm CM}$.

\begin{figure}[t]
\hspace*{-.25in}
\begin{center}
\epsfig{file=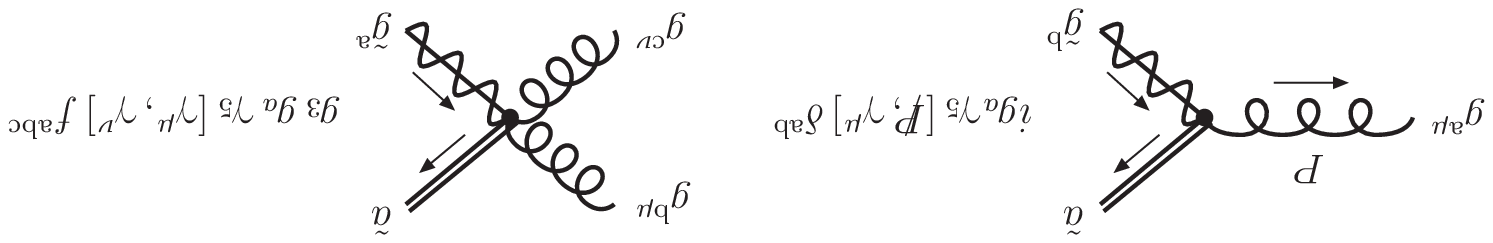,height=2.5cm,angle=180}
\end{center}
\hfill \caption[]{\sl Feynman rules used in the derivation of
$\tsq{ij}$ for the vertices with $\ax$. The arrows were drawn
according to the conventions of \cref{hk}.} \label{feyn}
\end{figure}

The computation of the various $\tsq{ij}$ can be realized applying
standard techniques \cite{hk}. We concentrate on the $\ax$
production processes which involve $SU(3)_{\rm C}$ interactions.
The Feynman rules for these interactions are indicated in
\Fref{feyn} and originate from the following Lagrangian term
\cite{axino, steffenaxino}:
\beq {\cal L}_{\ax\gl g}= i\,{{g_a}\over2}\,\bar{\ax}\,\gamma_5
\left[\gamma^{\mu},\gamma^{\nu}\right]\,\gl^{\rm a}\, G^{\rm
a}_{\mu\nu},~~\mbox{where}~~g_a=\frac{g^2_3}{32\pi^2 f_a}\,,
\label{Lagr} \eeq
$G^{\rm a}_{\mu\nu}=\partial_\mu g^{\rm a}_\nu-\partial_\nu g^{\rm
a}_\mu-g_3f^{\rm abc}g^{\rm b}_\mu g^{\rm c}_\mu$ is the gluon
field tensor, $f^{\rm abc}$ are the structure constants of the
$SU(3)_{\rm C}$ algebra with ${\rm a,~b,~c}=1,...,8$ whereas $\mu$
and $\nu$ are the usual spacetime indices. The residual Feynman
rules needed for our calculation are taken from \cref{pradler}. In
\Tref{tab1}, we list the Feynman diagrams included in the
calculation of $\tsq{ij}$. Note that in the non-relativistic
regime some of the processes ($gg$, $q\bar q$ and $gq$)
contributing to $\tsq{ij}$ in the relativistic regime
\cite{steffenaxino} are kinematically blocked and thus, are non
included in our computation.

In \Tref{tab2} we list analytical expressions for the non-averaged
squared amplitudes $g_ig_j\tsq{ij}$'s of the various scatterings
of $ij$ particles. Let us make some comments on the findings that
we present in this table:

\begin{itemize}

\item Although our numerical results for the $m_i$'s of \Eref{mi}
are practically independent on $m_{\ax}$ (since it is expected to
be significantly lower than $m_i$'s -- especially $m_{\gl}$ and
$m_{\sq}$), we explicitly display it in our formulas. Only gluons
and quarks are taken to be massless. Due to the smallness of
$m_{\ax}$ as compared to $m_i$'s, $C^{\rm LT}_{\ax}$ can be
approximated by an interpolated function of $T$, for every chosen
$m_i$'s. This is, in practice, a great simplification for our
numerical treatment.

\item The Feynman gauge has been used throughout our computation.
The substraction of the unphysical mode in the process $g\gl$ with
two external $g$'s can be obtained by employing appropriate
projection operators \cite{old} for the transverse polarization
states.

\item We present the complete expressions for all $g_ig_j\tsq{ij}$'s,
except for the case of $g\gl$, where $g_ig_j\tsq{ij}$ turns
out to be very lengthy; in this latter case, we display only the
important contributions which arise from the t-channel $\gl$
exchange and the interference terms between the $t$ and $s$ channel
and PI. We have checked that our results on the total squared
matrix elements for $\ax$ production agree nicely with those of
\cref{steffenaxino} in the relativistic limit (setting
$m_{\gl}=m_{\sq}=m_{\ax}=0$).

\item In the expressions we present, $g_ig_j\tsq{ij}$'s are
weighted with appropriate multiplicities following \cref{pradler}.
In particular, in the second column of \Tref{tab2} we arrange
factors due to the summation over all the color-triplets ($N_{\rm
F}=12$) -- taking into account the corresponding charge conjugate
multiplets which imply an extra factor of $2$. The factor $1/2$
occurs in the $\gl\gl$ process because of the identical particles
in the initial state. Factors arising from the summation over
color degrees of freedom and color indices, according to the
identities
\beq\label{idet}\sum_{\rm a,b,c}\left|f^{\rm
abc}\right|^2=N_3(N_3^2-1)~~\mbox{and}~~\sum_{a,I,J}\left|T^{\rm
a}_{IJ}\right|^2={1\over2}\left(N_3^2-1\right),\eeq
are included clearly in the expressions of the third column
($T^{\rm a}_{IJ}$ are the generators of the $SU(3)_{\rm C}$ group
and $N_3=3$).

\item The decay width of $\gl$ is evaluated employing the tree
level result of \cref{zerwas} -- possible next-to-leading order
corrections are subdominant. The result is checked to ensure that
it is consistent  with the output of the {\tt calcHEP} package
\cite{calchep}, and is the following:
\beq\Gamma_{\gl}=2N_{\rm F}{g_3(T)\over
32\pi}m_{\gl}\left(1-\left({m_{\sq}\over
m_{\gl}}\right)^2\right)^2\cdot\eeq

\item Although some integrals in \Eref{sig2} (especially the ones
corresponding to the processes $g\gl$ and $q\gl$) turn out to
converge slowly, we checked that our results are pretty stable
without the need to  introduce an effective mass for the
gluons propagators, as in the relativistic case \cite{Bolz, axino,
steffenaxino}.

\renewcommand{\arraystretch}{1.1}
\begin{table}[!t]
\begin{center}
\begin{tabular}{|c|c|c|}\hline
{\bf\textsc{Initial}}&{\bf\textsc{Final}} &{\bf\textsc{Interaction}}\\
{\bf\textsc{State}}&{\bf\textsc{State}} &{\bf\textsc{Channels}}\\
\hline \hline
$\gl^{\rm a}~\gl^{\rm b}$ & $g^{\rm c} ~\ax$ & $s(g^{\rm c}),~t(g^{\rm b}),~u(g^{\rm a})$ \\
$\sq_I~\gl^{\rm a}$ & $\sq_J~ \ax$ & $t(g^{\rm a})$  \\
$g^{\rm a} ~\gl^{\rm b}$ & $\gl^{\rm c} ~\ax$ &  $ s(\gl^{\rm c}),~t(g^{\rm b}),~u(\gl^{\rm a}),~{\rm PI}$\\
$q_I~\gl^{\rm a} $ & $q_J ~\ax$ & $t(g^{\rm a})$ \\
$\sq_I ~\sq^*_J$ & $\gl^{\rm a} ~\ax$ & $s(g^{\rm a})$ \\
$\sq_I ~g^{\rm a}$ & $q_J ~\ax$ & $t(\gl^{\rm a})$ \\
$\sq^*_I ~q_J$ & $g^{\rm a} ~\ax$ & $s(\gl^{\rm a})$ \\
\hline
\end{tabular}
\end{center}
\caption{\sl\ftn The Feynman diagrams contributing to $\sigv{ij}$
with \ax\ in the final state. The exchanged particles are
indicated for each relevant pair of initial and final states. The
symbols $s(i)$, $t(i)$ and $u(i)$ denote tree-graphs in which the
particle $i$ is exchanged in the s-, t- or u-channel and PI stands
for ``point interaction''. The superscripts are $SU(3)$ color
indices, whereas the subscript $I$ and $J$ are family indices.
\label{tab1}}
\end{table}

\item In our numerical computation and with the $m_i$'s listed in
\Eref{mi}, we do not include the contributions of the processes
$\sq\sq^*$ and $\sq g$ since they are, in general, negligible. The
major contributions to $\tsq{ij}$ come from $\gl \gl$ and $\sq
\gl$ for $T\simeq\Ts$ and from $\sq q$ for $T\ll\Ts$. The
processes $g\gl$ and $q\gl$ give in general important
contributions which cannot be neglected. Note, however, that the
contribution of the process $\sq^* q$ could also be enhanced, if
$m_{\gl}\simeq m_{\sq}$, since we would have a resonance in the
$\sq-q$ annihilation via an $s$-channel exchange of a $\gl$.
However, we consider that the choice $m_{\gl}\simeq m_{\sq}$ would
be a rather ugly tuning of our free parameters and thus, we opted
to use the $m_i$'s in \Eref{mi} in order to demonstrate our
findings.

\end{itemize}

\renewcommand{\arraystretch}{1.5}
\begin{table}[!t]
\begin{center}
\begin{tabular}{|c||c|l|}\hline
{$ij$}&\multicolumn{2}{|c|}{$g_ig_j\tsq{ij}/g_a^2g_3^2$}\\
\hline\hline
$\gl\gl$&${1\over2}$&\hspace*{-2.6cm}\begin{minipage}[h]{13cm}
\vspace*{-.2cm}\begin{eqnarray*}&& {32N_3(N_3+1)\over
(stu)^2}\Big[2 m_{\gl}^2 (m_{\ax}^2 - m_{\gl}^2)^2 s^2 t^2 + s t
\Big(4 m_{\gl}^3 (m_{\gl} - m_{\ax}) (m_{\gl} + m_{\ax})^2 s
\\&& -\,(m_{\gl} - m_{\ax})\left(2 m_{\gl}^2 (2 m_{\gl} - m_{\ax})
(m_{\gl} + m_{\ax})^2 + (m_{\gl} - m_{\ax}) (3 m_{\gl}^2 + 2
m_{\gl} m_{\ax}\right.
\\ && \left.+\, m_{\ax}^2) s\right) t + 2 s (3 m_{\gl}^2 - 2 m_{\gl} m_{\ax} + s) t^2\Big) u
-\Big(2 m_{\gl}^2 (m_{\gl}^2 - m_{\ax}^2)^2 s^2+ st\left( (7
m_{\gl}^4\right.
\\&& \left. +\, m_{\ax}^4) s + 2 m_{\gl}^2 (m_{\gl} + m_{\ax})^2 (2
m_{\gl}^2 - 3 m_{\gl} m_{\ax} + m_{\ax}^2)\right)+ \left(2
m_{\gl}^2 (m_{\gl}^2 - m_{\ax}^2)^2\right.
\\&&  \left.+\, (3 m_{\gl}^4 + 4
m_{\gl}^3 m_{\ax} + m_{\ax}^4) s - (m_{\gl} - m_{\ax}) (5 m_{\gl}
+ 3 m_{\ax}) s^2 - 2 s^3\right) t^2 - 2 s^2 t^3\Big) u^2
\\&& +\, 2 s t \left(-s t + 3 m_{\gl}^2 (s + t) + m_{\ax}^2 (s +
t)\right) u^3 - 2 s t (s + t) u^4\Big]\\[-.2cm]
\end{eqnarray*}
\end{minipage}\\ \hline
$\sq\gl$& $2N_{\rm F}$&\hspace*{-2.6cm}
\begin{minipage}{13cm}\vspace*{-.2cm}\begin{eqnarray*}&& {4\over t^2}(N_3^2-1)\Big(
2 m_{\gl}^4 m_{\ax}^2 t\,(t-4 m_{\sq}^2)-4 (m_{\gl}^2+m_{\ax}^2)
m_{\sq}^2 \\ && - t\, (t + 2 u-2 m_{\sq}^2)+ m_{\ax}^2 t (t + 4 u)
+ m_{\gl}^2\left(4 m_{\ax}^2  (2
m_{\sq}^2 - t) + t (t + 4 u)\right)\Big)\\[-.2cm]\end{eqnarray*}
\end{minipage}\\ \hline
$g\gl$& $1$&\hspace*{-2.6cm}
\begin{minipage}[h]{13cm}\vspace*{-.2cm}\begin{eqnarray*}&& N_3(N_3-1)\Big[
 {8\,(m_{\gl}^2-s)\over s^2t\,\left((s-m_{\gl}^2)^2+\Gamma^2_{\gl}m_{\gl}^2\right)}\Big(4s m_{\gl}^2
\left(m^2_{\gl} - m^2_{\ax}\right)
\left(s-m_{\ax}^2\right)(m_{\gl}^2 - 2 m_{\ax}^2
\\&& + s)+\left(m_{\gl}^6 (m_{\ax}^2 - 2 s) - 4
m_{\ax}^4 s^2 + m_{\gl}^5 (6 m_{\ax}^3 - 4 m_{\ax} s) + m_{\gl}^4
(m_{\ax}^4 - 21 m_{\ax}^2 s + 12 s^2)\right.
\\&&\left.-2 m_{\gl}^3 (2 m_{\ax}^5 + 5 m_{\ax}^3 s - 6 m_{\ax} s^2) +
m_{\gl}^2 (6 s^3-2 m_{\ax}^6 + 27 m_{\ax}^4 s - 18 m_{\ax}^2
s^2)\right) t
\\&& -\left(m_{\gl}^2 m_{\ax}^2 (3 m_{\gl}^2 - 6 m_{\gl} m_{\ax} - 8
m_{\ax}^2) - (10 m_{\gl}^4 + 28 m_{\gl}^3 m_{\ax} + m_{\gl}^2
m_{\ax}^2 - 2 m_{\ax}^4) s\right.+
\\&& \left. 2 (m_{\gl} + m_{\ax}) (3 m_{\gl} + m_{\ax}) s^2\right) t^2 - 2
t^3\left(2 m_{\gl} m_{\ax} s + 2 m_{\gl}^2 (m_{\ax}^2 + 2 s) - s
(m_{\ax}^2 + 2 s)\right)\Big)
\\ && +{4\over s^2t} \Big(16 m_{\gl}^6 s + 16 m_{\ax}^6 s + 2 m_{\gl}^3
m_{\ax} t (8 t-3 m_{\ax}^2 + 12 s) - 4 m_{\ax}^4 (6 s^2 + 3 s t -
2 t^2)
\\ &&+ m_{\gl}^4 \left(8 t^2-4 (m_{\ax}^4 + 3 m_{\ax}^2 s + 6 s^2) +
(m_{\ax}^2 -12 s)\ t \right) + 4 m_{\ax}^2 (4 s^3 + 4 s^2 t - 5 s
t^2
\\&&- 4 t^3)-4 (2 s + t) (2 s^3 + 3 s^2 t - 2 t^3) + 8t m_{\gl} m_{\ax}
\left(m_{\ax}^2 (3 s + 2 t) -s^2 - 4 s t - 2 t^2 \right)
\\&&+ m_{\gl}^2 \left(m_{\ax}^4 (t-12 s) + m_{\ax}^2 (28 s^2 + 4 s
t + 19 t^2) + 4 (4 s^3 + 4 s^2 t - 5 s t^2 - 4
t^3)\right)\Big)\Big]\\[-.2cm]
\end{eqnarray*}
\end{minipage}\\ \hline
$q\gl$& $2N_{\rm F}$&\hspace*{-2.6cm}
\begin{minipage}{13.0cm}\vspace*{-.2cm}
\begin{eqnarray*}{{8}(N_3^2-1)}\Big((m_{\gl}^2+m_{\ax}^2)(2 s +
t)-m_{\gl}^4-m_{\ax}^4-2 m_{\gl} m_{\ax} t-2 s (s + t)
\Big)/t\\[-.2cm]\end{eqnarray*}
\end{minipage}\\ \hline
$\sq\sq^*$& $N_{\rm F}$&
\begin{minipage}[h]{13cm} The same as for the process $\sq\gl$ but with the
opposite sign and $s~\leftrightarrow~t$
\end{minipage}\\ \hline
$\sq g$& $2N_{\rm F}$&\hspace*{-2.6cm}
\begin{minipage}[h]{13.0cm}\vspace*{-.2cm}
\begin{eqnarray*}{8}(N_3^2-1)(t-m_{\ax}^2) (s t -m_{\ax}^2 m_{\sq}^2 + m_{\gl}^2
u)/(t-m_{\gl}^2 )^2\\[-.2cm]\end{eqnarray*}
\end{minipage}\\ \hline\renewcommand{\arraystretch}{2.5}
$\sq^* q$& $2N_{\rm F}$&\hspace*{-2.6cm}
\begin{minipage}[h]{13cm}\vspace*{-.2cm}
\begin{eqnarray*}{8}(N_3^2-1)(s-m_{\ax}^2) (s t -m_{\ax}^2 m_{\sq}^2 +  m_{\gl}^2
u)/\left((s-m_{\gl}^2
)^2+\Gamma_{\gl}^2m_{\gl}^2\right)\\[-.2cm]\end{eqnarray*}
\end{minipage}\\ \hline
\end{tabular}\end{center}
\caption{\sl\ftn The quantities $g_ig_j\tsq{ij}$ for $\ax$
production from scatterings of the $ij$ particles in the
non-relativistic regime. The results are summed over spins,
generation and color indices, in the initial and final state. The
Mandelstam variables are defined as $s=(p_1 + p_2)^2, t=(p_1 -
k_1)^2$ and $u=(p_1 - k_2)^2$ where the four-momenta $p_1$, $p_2$,
$k_1$, and $k_2$ are associated with the particles in the order in
which they are written down in the columns ``initial state'' and
``final state'' of \Tref{tab1}.\\[.4cm]}\label{tab2}
\end{table}

\section*{References}

\rhead[\fancyplain{}{ \bf \thepage}]{\fancyplain{}{\sc
Quintessential Kination and Thermal Production of Gravitinos and
Axinos }} \lhead[\fancyplain{}{\sc References }]{\fancyplain{}{\bf
\thepage}} \cfoot{}

\end{document}